%% file: Koopman_Vehicles_Survey_rev_black.tex
\documentclass{IEEEtran}
\usepackage{cite}
\usepackage{amsmath,amssymb,amsfonts}
\usepackage{graphicx}
\usepackage{textcomp}
\usepackage{float}           
\newtheorem{remark}{Remark}  

\begin{document}
	\title{Vehicular Applications of Koopman Operator Theory -- A Survey}
	\author{W. Manzoor, S. Rawashdeh, and A. Mohammadi}
	
	\maketitle
	
	\begin{abstract}
		Koopman operator theory has proven to be a promising approach to nonlinear system identification and global linearization. For nearly a century, there had been no efficient means of calculating the Koopman operator for applied engineering purposes. The introduction of a recent computationally efficient method in the context of fluid dynamics, which is based on the system dynamics decomposition to a set of normal modes in descending order,  has overcome this long-lasting computational obstacle. The purely data-driven nature of Koopman operators holds the promise of capturing unknown and complex dynamics for reduced-order model generation and system identification, through which the rich machinery of linear control techniques can be utilized. Given the ongoing development of this research area and the many existing open problems in the fields of smart mobility and vehicle engineering, a survey of techniques and open challenges of applying Koopman operator theory to this vibrant area is warranted. This review focuses on the various solutions of the Koopman operator which have emerged in recent years, particularly those focusing on mobility applications, ranging from characterization and component-level control operations to vehicle performance and fleet management. Moreover, this comprehensive review of over 100 research papers highlights  the breadth of ways Koopman operator theory has been applied to various vehicular applications with a detailed categorization  of the applied Koopman operator-based algorithm type. Furthermore, this review paper discusses theoretical aspects of Koopman operator theory that have been largely neglected by the smart mobility and vehicle engineering community and yet have large potential for contributing to solving open problems in these areas. 
	\end{abstract}

\input{./intro}

\input{./genTax}

\input{./briefOver}

\input{./vehicAppli}

\input{./vehicAppli2}

\input{./Conclusion}

\section*{Acknowledgments}
We would like to acknowledge our fruitful discussion with Professor Igor Mezi\'{c} regarding the historical developments associated with Koopman operator theory and its applications.  This work was partially funded by the Michigan Space Grant Consortium, NASA: Award Number~\#NNX15AJ20H. 

\bibliographystyle{IEEEtran}
\bibliography{Koopman_Vehicles_Survey_rev_black}

\end{document}

%% file: intro.tex
\section{Introduction}
\label{sec:intro}

\IEEEPARstart{K}{oopman} operator theory is named after Bernard Koopman, who in the 1930s proved the premise that linear transformations of nonlinear dynamical systems exist when represented in Hilbert [function] space \cite{koopman:1931}. Historically, determining a Koopman-invariant subspace was accomplished by trial and error despite being unsuccessful for most dynamical systems. The enabling  engine for  modern-day data-driven applications of the Koopman operator theory are due to the profound insights on geometrical and statistical properties of dynamical systems in the Ph.D. dissertation~\cite{mezic1994geometrical} and the foundational line of work~\cite{mezic2005spectral,mezic2004comparison} on harmonic  analysis of the  Koopman operator.  Mezi\'{c}'s pioneering work~\cite{mezic1994geometrical,mezic2005spectral,mezic2004comparison} along  with   the  computational breakthroughs  based on singular value decomposition (SVD) enabled approximation of the Koopman operator from large amounts of data without relying on the pseudo-inversion of large non-square matrices. It is remarked that \emph{the first modern-day engineering application} of Koopman operator theory has been due to Mezi\'{c} and Banaszuk~\cite{mezic2004comparison} for model parameter identification in combustion rigs. 

\begin{figure}[t]
\centering
\includegraphics[width=0.4\textwidth]{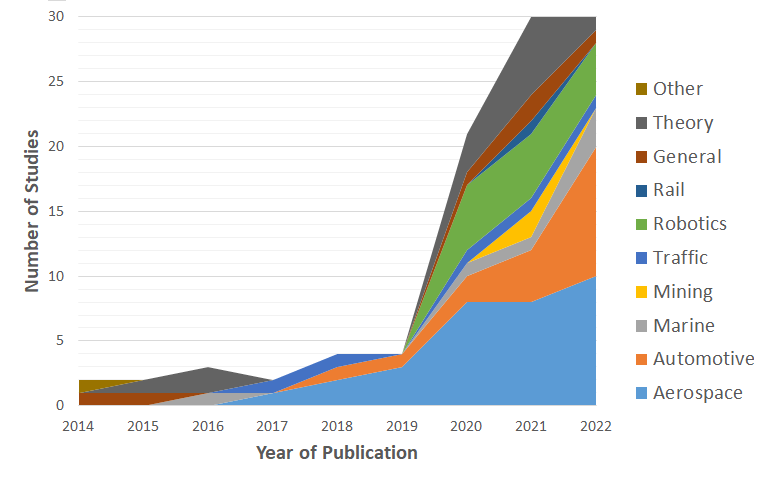}
\caption{\small Publication timeline of the surveyed literature. The number of studies incorporating Koopman operator-based methods in smart mobility and vehicle engineering applications has been increasing nearly exponentially over an eight-year span thus far.}
\label{fig:Timeline_area}   
\end{figure}

As demonstrated in Figure~\ref{fig:Timeline_area}, \emph{the first vehicular applications} started to emerge only six years after the computational breakthrough due to the Dynamic Mode Decomposition (DMD) technique was achieved in 2008~\cite{schmid:2008}. Among vehicular applications alone, it is evident from Figure~\ref{fig:Timeline_area} that the number of studies incorporating Koopman operator-based methods has been increasing nearly exponentially over an eight-year time span since 2014. However, with a mere maximum of 30 such studies published so far in a given year, the likely trend is further exponential growth as more researchers become aware of the associated algorithms, and as the applicability of such algorithms simultaneously evolves. 

\noindent \textbf{Features of the Current Survey Paper.} The purely data-driven nature of Koopman operators holds the promise of capturing unknown and complex dynamics for reduced-order model generation and system identification, through which the rich machinery of linear control techniques can be utilized. The emergent nature of the smart mobility and vehicular-related applications, where  the Koopman operator  in each particular application needs to be approximated, implies that the development of various Koopman operator approximation  algorithms is expected to grow along with the vehicular problems they aim to solve.  Given the ongoing development of this research area and the many existing open problems in the fields of smart mobility and vehicle engineering, a survey of techniques and open challenges of applying Koopman operator theory to this vibrant area is warranted.  To the best of our knowledge, this survey paper is the \emph{first of its kind} reviewing the applications of Koopman operator theory within a focused research area, namely, smart mobility and vehicle engineering applications. A \emph{notable feature} of our survey paper is reviewing and categorizing the results of over 100 research papers based on both application and algorithm type  (see Table~\ref{tab1} and Section~\ref{sec:vehicApp}) that are concerned with the applications of Koopman operator theory to the field of smart mobility and vehicular engineering. Such a \emph{comprehensive and  detailed categorization} will be beneficial to the research practitioners working in the field.  Furthermore, this review paper discusses theoretical aspects of Koopman operator theory that have been largely neglected by the smart mobility and vehicle engineering community and yet have large potential for contributing to solving open problems in these areas. Additionally, our survey paper seeks to \emph{identify gaps} in the smart mobility and vehicle engineering research where new and existing Koopman operator-based methods have the potential to further develop and address unsolved problems  potentially benefiting from the perspectives of nonlinear system identification, control, global linearization, and the predictive powers that Koopman operator theory has to offer (see, e.g., Remarks~\ref{remGap1}--\ref{remGap6}). 

\input{./table1}

The rest of this paper is organized as follows.  After presenting the relevant taxonomy in  Section~\ref{sec:genTax}, we provide a brief overview  the basic underpinnings of the Koopman operator theory in Section~\ref{sec:briefOver}. The literature review with categorized vehicular applications is presented in  Section~\ref{sec:vehicApp}, where each subsection concludes with a list of open research questions for the application of Koopman operator-based methods in terms of vehicle types not encountered in the literature.   Other relevant applications, which are not explicitly vehicular in nature, and theoretical/algorithmic variations are reviewed in  Section~\ref{sec:vehicApp2}.   

\begin{remark}
A survey paper on variants of DMD authored by Chen \emph{et al.}~\cite{chen:2012} had been published in 2012. However, being conducted a decade ago, it was well before the emergence of Koopman operator vehicular applications in the literature. Nevertheless, a very interesting discussion on the optimal application of DMD can be found in that paper. Another influential review in the field of Koopman operator theory is due to Budi{\v{s}}i{\'c} \emph{et al.}~\cite{budivsic2012applied}.   Over the course of developing this paper, Schmid~\cite{schmid:2022}, who is the pioneer of the original mode decomposition method, has also published his own survey on the variants of his method found in the literature. However, Schmid's survey is not focused on the specific topic of smart mobility and vehicle engineering applications. Finally, we remark that while Schmid's survey paper is focused on DMD and its variants, a complete review on Koopman operator methods in fluid mechanics due to Mezi\'{c} has been presented in~\cite{mezic2013analysis}.
\label{rem:features}
\end{remark}

%% file: table1.tex
\begin{table*}[!t]
	\caption{Summary of studies organized by the utilized Koopman operator-based algorithm}
	\begin{tabular}{c | c | c | c }
		\hline
		\centering\textbf{Algorithm} & \textbf{Studies$^\ddag$} & 
		\centering\textbf{Algorithm} & \textbf{Studies$^\ddag$} \\  \hline
		\begin{minipage}{0.33\textwidth}
		\vspace{1ex}
		Direct Method of Selecting Koopman Observables (\textbf{DMa})
		\vspace{1ex}
		\end{minipage}
	    &
	    \begin{minipage}{0.1\textwidth}
		 \cite{chen:2020}\cite{cibulka:2020}  
		\end{minipage}
 		& 
 		\begin{minipage}{0.33\textwidth}
 			\vspace{1ex}
 			Projection Methods onto Selected/Derived/Required Basis, including use of Principle Orthogonal Decomposition (\textbf{POD})
 			\vspace{1ex}
 		\end{minipage}
 		&
 		\begin{minipage}{0.1\textwidth}
 			\cite{hofmann:2022}\cite{qian:2020}\cite{renganathan:2018}\\
 			\cite{otto:2022}  
 		\end{minipage}
		\\ \hline
		%
		\begin{minipage}{0.33\textwidth}
		\vspace{1ex}
		Direct Methods using Frobenius--Perron Method or [Factorization of PDEs for] Adjoint Koopman Operator, 
		Method of Characteristics, Legendre polynomials or Schur Decomposition (\textbf{DMb})
		\vspace{1ex}
		\end{minipage}
		& 
		\begin{minipage}{0.1\textwidth}
		\cite{leonard:2019}\cite{schmid:2022}\cite{gutow:2020}\cite{meyers:2019}\\
		 \cite{sinha:2016}\cite{servadio:2022b}\cite{salam:2022}\cite{arnas:2022}  
		\end{minipage}
		&
		\begin{minipage}{0.33\textwidth}
			\vspace{1ex}
			Dynamic Mode Decomposition, excluding where DMD is compared against modified methods (\textbf{DMD})
			\vspace{1ex}
		\end{minipage}
		& 
		\begin{minipage}{0.1\textwidth}
			\cite{cohen:2021}\cite{schmid:2022}\cite{wu:2021}\\
			\cite{ling:2020}\cite{ling:2018} \cite{boudali:2017}\\
			\cite{hanke:2018}\cite{le:2015}\cite{vasisht:2020} \\
			\cite{journell:2020}\cite{luong:2021}\cite{larusson:2014}\\
			\cite{avila:2021} \cite{manzoor:2022b}  
		\end{minipage}
		\\ \hline
		\begin{minipage}{0.33\textwidth}
			\vspace{1ex}
		Direct Methods including Error Minimization (e.g., using Taylor Series Error Bounds), Minimization of Frobenius Norm through Pseudo-inversion of Data Matrix or other Optimization (\textbf{DMc})
		\vspace{1ex}
		\end{minipage}
		& 
		\begin{minipage}{0.1\textwidth}
		\cite{mamakoukas:2021}\cite{zinage:2021}\cite{sinha:2022}  
	    \end{minipage}
		&
		\begin{minipage}{0.33\textwidth}
			\vspace{1ex}
			Extended DMD (\textbf{EDMD}) or slightly modified EDMD, e.g., Naturally Structured DMD (\textbf{NSDMD})
			\vspace{1ex}
		\end{minipage}
		& 
		\begin{minipage}{0.1\textwidth}
			\cite{zhu:2022}\cite{berrueta:2020}\cite{schmid:2022}\\
			\cite{cibulka:2019}\cite{vsvec:2021a}\cite{mi:2022}\\
			\cite{vsvec:2021b}\cite{zinage:2022a}\cite{buzhardt:2022}\\
			\cite{lehmberg:2021}\cite{bhandari:2021}\cite{li:2021}\\
			\cite{williams:2016}\cite{calderon:2022}\cite{bruder:2021}\\
			\cite{shi:2021}\cite{chen:2021}\cite{williams:2015}\\
			\cite{moyalan:2022}\cite{broad:2020}\cite{broad:2018}\\
			\cite{pan:2021}
		\end{minipage}
		\\ \hline
		\begin{minipage}{0.33\textwidth}
			\vspace{1ex}
		Duel Faceted Linearization (\textbf{DFL}) 
		\vspace{1ex}
		\end{minipage}
		& 
		\begin{minipage}{0.1\textwidth}
		\cite{sotiropoulos:2021}  
	    \end{minipage}
		&
		\begin{minipage}{0.33\textwidth}
			\vspace{1ex}
			Weighted Online EDMD (\textbf{WOEDMD})
			\vspace{1ex}
		\end{minipage}
		& 
		\begin{minipage}{0.1\textwidth}
			\cite{guo:2022}  
		\end{minipage}
		\\ \hline
		\begin{minipage}{0.33\textwidth}
			\vspace{1ex}
		Bilinear Koopman Realization (\textbf{BKR})
			\vspace{1ex}
     	\end{minipage}
		& 
		\begin{minipage}{0.1\textwidth}
		\cite{yu:2022}  
   		\end{minipage}
		&
		\begin{minipage}{0.33\textwidth}
			\vspace{1ex}
			DMD with Control (\textbf{DMDc}) 
			\vspace{1ex}
		\end{minipage}
		& 
		\begin{minipage}{0.1\textwidth}
			\cite{jin:2020}\cite{manzoor:2022b} 
		\end{minipage}
		\\ \hline
		%
		\begin{minipage}{0.33\textwidth}
			\vspace{1ex}
		 Robust Koopman MPC (\textbf{RK-MPC}) or Robust Tube-based MPC with Koopman (\textbf{r-KMPC}) 
		 \vspace{1ex}
		\end{minipage}
		& 
		\begin{minipage}{0.1\textwidth}
		\cite{mamakoukas:2022}\cite{zhang:2022}  
		\end{minipage}
		&
		\begin{minipage}{0.33\textwidth}
			\vspace{1ex}
			Koopman Eigenfunction Extended Dynamic Mode Decomposition (\textbf{KEEDMD}) or modified KEEDMD
			\vspace{1ex} 
		\end{minipage}
		& 
		\begin{minipage}{0.1\textwidth}
			\cite{folkestad:2020a}\cite{folkestad:2020b}  
		\end{minipage}
		\\ \hline
		\begin{minipage}{0.33\textwidth}
			\vspace{1ex}
			Koopman Mode Decomposition (\textbf{KMD}) and/or determination of Koopman modes only (not Koopman operator), including through Arnoldi iteration (Note: KMD was first discovered in~\cite{mezic2005spectral}.)
	\vspace{1ex}	
	\end{minipage}
	&
	    \begin{minipage}{0.1\textwidth}
		  \cite{mezic2005spectral}\cite{avila:2017}\cite{avila2020data}\cite{hogg:2020}\cite{canham:2017}  
	\end{minipage}
    &
    \begin{minipage}{0.33\textwidth}
    	\vspace{1ex}
    	Rescaled DMD (\textbf{rDMD}) 
    	\vspace{1ex}
    \end{minipage}
    &
    \begin{minipage}{0.1\textwidth}
    	\cite{cohen:2021}  
    \end{minipage}
    \\ \hline
		%
		
		\begin{minipage}{0.33\textwidth}
			\vspace{1ex}
		Koopman Map Inversion (\textbf{KMI})
			\vspace{1ex}
		\end{minipage}
	& 
		\begin{minipage}{0.1\textwidth}
				\cite{servadio:2022a}  
		\end{minipage}
	& 
	\begin{minipage}{0.33\textwidth}
		\vspace{1ex}
		Debiased DMD, Forward/Backward DMD (\textbf{FB-DMD})
		\vspace{1ex}
	\end{minipage}
	& 
	\begin{minipage}{0.1\textwidth}
		\cite{schmid:2022}
	\end{minipage}	
	\\ \hline
		\begin{minipage}{0.33\textwidth}
		\vspace{1ex}
		Debiased DMD, Forward/Backward DMD 
		\vspace{1ex}
	\end{minipage}
	&
	\begin{minipage}{0.1\textwidth}
		\cite{schmid:2022} 
	\end{minipage}
	& 
	\begin{minipage}{0.33\textwidth}
		\vspace{1ex}
		Deep Koopman, Deep Direct Koopman, or other general Artificial Neural Network (ANN) Scheme (\textbf{DDK-ANN})
		\vspace{1ex}
	\end{minipage}
	& 
	\begin{minipage}{0.1\textwidth}
		\vspace{0.5ex}
		\cite{xiao:2021_i13}\cite{xiao:2022}\cite{wang:2021}\\
		\cite{han:2020} \cite{wang:2022}\cite{bakker:2020}\\
		\cite{zhan:2022}\cite{maksakov:2020} \cite{jin:2020} 
		\vspace{0.5ex}
	\end{minipage}
	\\ \hline
	%
	%
	\begin{minipage}{0.33\textwidth}
		\vspace{1ex}
		Multi-resolution DMD (\textbf{mrDMD}) 
		\vspace{1ex}
	\end{minipage}
	& 
	\begin{minipage}{0.1\textwidth}
		\cite{schmid:2022}\cite{sullivan:2021} 
	\end{minipage}
	& 
	\begin{minipage}{0.33\textwidth}
		\vspace{1ex}
		Neural Koopman Lyapunov Control [ANN] (\textbf{NKLC})  
		\vspace{1ex}
	\end{minipage}
	& 
	\begin{minipage}{0.1\textwidth}
		\cite{zinage:2022b}  
	\end{minipage}
	\\ \hline
	\begin{minipage}{0.33\textwidth}
		\vspace{1ex}
		HAVOK, Hankel DMD (partial or full-predictive)  (\textbf{HAVOC}) (Note: Hankel DMD was first introduced by Arbabi and Mezi\'{c}\cite{arbabi2017ergodic}.)
		\vspace{1ex}
	\end{minipage}
	& 
	\begin{minipage}{0.1\textwidth}
		\cite{arbabi2017ergodic}\cite{schmid:2022}\cite{ali:2021}\cite{haggerty:2020}\\
		\cite{ma:2022}\cite{manzoor:2022a}
	\end{minipage}
	& 
	\begin{minipage}{0.33\textwidth}
		\vspace{1ex}
		Deterministic/Convolutional Koopman Network (\textbf{DCKNet}, \textbf{CKNet}), Deep Convolutional  Koopman Network (\textbf{DKN})   
		\vspace{1ex}
	\end{minipage}
	&
	\begin{minipage}{0.1\textwidth}
		\cite{xiao:2021_i31}\cite{leask:2021}
	\end{minipage}
	\\ \hline
	\begin{minipage}{0.33\textwidth}
		\vspace{1ex}
		Higher-order DMD (\textbf{HODMD}) 
		\vspace{1ex}
	\end{minipage}
	& 
	\begin{minipage}{0.1\textwidth}
		\cite{schmid:2022}\cite{le:2019} 
	\end{minipage}
	& 
	\begin{minipage}{0.33\textwidth}
		\vspace{1ex}
		Deep Learning with Recurrent Neural Network  (\textbf{DLRN})
		\vspace{1ex}
	\end{minipage}
	& 
	\begin{minipage}{0.1\textwidth}
		\cite{bieker:2020} 
	\end{minipage}
	
	\\ \hline
	\begin{minipage}{0.33\textwidth}
		\vspace{1ex}
		Physics-based Higher-order EDMD (\textbf{PhysEDMD})
		\vspace{1ex}
	\end{minipage}
	& 
	\begin{minipage}{0.1\textwidth}
		\cite{castano:2020}
	\end{minipage}
	& 
	\begin{minipage}{0.33\textwidth}
		\vspace{1ex}
		Direct Koopman Reinforcement Learning (\textbf{DKRL}) 
		\vspace{1ex}
	\end{minipage}
	&
	\begin{minipage}{0.1\textwidth}
		\cite{songy:2021}
	\end{minipage}
	\\ \hline
	%
	\begin{minipage}{0.33\textwidth}
		\vspace{1ex}
		Bilinear EDMD (\textbf{bEDMD}) 
		\vspace{1ex}
	\end{minipage}
	& 
	\begin{minipage}{0.1\textwidth}
		\cite{folkestad:2021}   
	\end{minipage}
	& 
	\begin{minipage}{0.33\textwidth}
		\vspace{1ex}
		Split Koopman Autoencoder (\textbf{SKA})  
		\vspace{1ex}
	\end{minipage}
	& 
	\begin{minipage}{0.1\textwidth}
		\cite{girgis:2021}
	\end{minipage}
	\\ \hline
	\begin{minipage}{0.33\textwidth}
		\vspace{1ex}
		Constant Energy Multiscale DMD (\textbf{CEM-DMD})
		\vspace{1ex}
	\end{minipage}
	&
	\begin{minipage}{0.1\textwidth}
		\cite{li:2021}
	\end{minipage}
	& 
	\begin{minipage}{0.33\textwidth}
		\vspace{1ex}
		Stepwise Akaike Information Criteria (\textbf{SAIC})  
		\vspace{1ex}
	\end{minipage}
	&
	\begin{minipage}{0.1\textwidth}
		\cite{matpan:2021}
	\end{minipage}
	\\ \hline
	
	\begin{minipage}{0.33\textwidth}
		\vspace{1ex}
		Exact DMD, ANN-based Exact DMD  (\textbf{ANN-EDMD})
		\vspace{1ex}
	\end{minipage}
	& 
	\begin{minipage}{0.1\textwidth}
		\cite{schmid:2022}\cite{li:2021}
	\end{minipage}
	& 
	\begin{minipage}{0.33\textwidth}
		\vspace{1ex}
		EDMD with Autodidact Stiffness Learning (\textbf{ASL-EDMD})
		\vspace{1ex}
	\end{minipage}
	& 
	\begin{minipage}{0.1\textwidth}
		\cite{goyal:2022}  
	\end{minipage}
	\\ \hline
	\begin{minipage}{0.33\textwidth}
		\vspace{1ex}
		\textbf{LIR-DMD} 
		\vspace{1ex}
	\end{minipage}
	&
	\begin{minipage}{0.1\textwidth}
		\cite{li:2021}
	\end{minipage}
	& 
	\begin{minipage}{0.33\textwidth}
		\vspace{1ex}
		Sparse Identification of Nonlinear Dynamics (\textbf{SINDy}) 
		\vspace{1ex}
	\end{minipage}
	&
	\begin{minipage}{0.1\textwidth}
		\cite{brunton:2016}\cite{matpan:2021} 
	\end{minipage}
	\\ \hline
	%
	\begin{minipage}{0.33\textwidth}
		\vspace{1ex}
		Time Delay DMD (\textbf{TD-DMD}) 
		\vspace{1ex}
	\end{minipage}
	& 
	\begin{minipage}{0.1\textwidth}
		\cite{pan:2021} 
	\end{minipage}
	& 
	\begin{minipage}{0.33\textwidth}
		\vspace{1ex}
		Stochastic Koopman Operator (\textbf{SKO}), or Stochastic Adversarial Koopman Operator with Auxillary Neural Network, Gaussian Process-based Koopman Operator, or other Stochastic Methods   
		\vspace{1ex}
	\end{minipage}
	& 
	\begin{minipage}{0.1\textwidth}
		\cite{balakrishnan:2021}\cite{yingzhao:2020}\\\cite{bujorianu:2021}
		\cite{williams:2015}
	\end{minipage}
	\\ \hline
	\begin{minipage}{0.33\textwidth}
		\vspace{1ex}
		Kernel DMD (\textbf{KDMD})  
		\vspace{1ex}
	\end{minipage}
	& 
	\begin{minipage}{0.1\textwidth}
		\cite{pan:2021}
	\end{minipage}
	& 
	\begin{minipage}{0.33\textwidth}
		\vspace{1ex}
		Sparcity Promoting DMD (\textbf{spDMD})   
		\vspace{1ex}
	\end{minipage}
	& 
	\begin{minipage}{0.1\textwidth}
		\cite{pan:2021}
	\end{minipage}
	\\ \hline
	\end{tabular}
    \newline
	\footnotesize {$^\ddag$ Detailed descriptions are provided in the text.}
	\label{tab1}
\end{table*}

%% file: genTax.tex
\section{General Taxonomy and Vehicle Type Categorization}
\label{sec:genTax}

For the purposes of this survey, we generally  consider a \emph{vehicle} to be any man-made instrument that can carry a payload, including occupants, equipment, sensors or any other items. In some contexts, the item may be its own presence. Most generally, vehicle has also been defined as any mechanized equipment. All these ideas have been captured in Merriam-Webster's formal definitions \cite{merriamwebster:vehicle}. In this paper, the physical type of vehicle according to Merriam-Webster  is considered. This physical type includes systems and processes associated with automotive, aerospace, marine, rail, robotic, and subterranean classes of vehicles, along with some interfacing or noteworthy classes such as biolocomotion, communication, and traffic management, amongst others.

The \emph{motivation} to focus on vehicular applications stems from the fact that many processes and subsystems are not easily modeled to a sufficient level of fidelity and/or are subject to a significant level of noise/disturbances.  Such conditions pose limitations on the performance, control and overall utilization of the processes and subsystems that transduce energy into motion. Vehicles of different types, such as aircraft and automobiles, may further share common subsystems (e.g., combustion chamber or pump) or types of maneuvers (e.g., braking or collision avoidance).  Consequently, from a dynamics and controls standpoint, it is sufficient to maintain the scope of this study to include all major vehicle categories.

\begin{figure*}[t]
	\centering
	\begin{minipage}{0.3\textwidth}
		\includegraphics[width=0.95\textwidth]{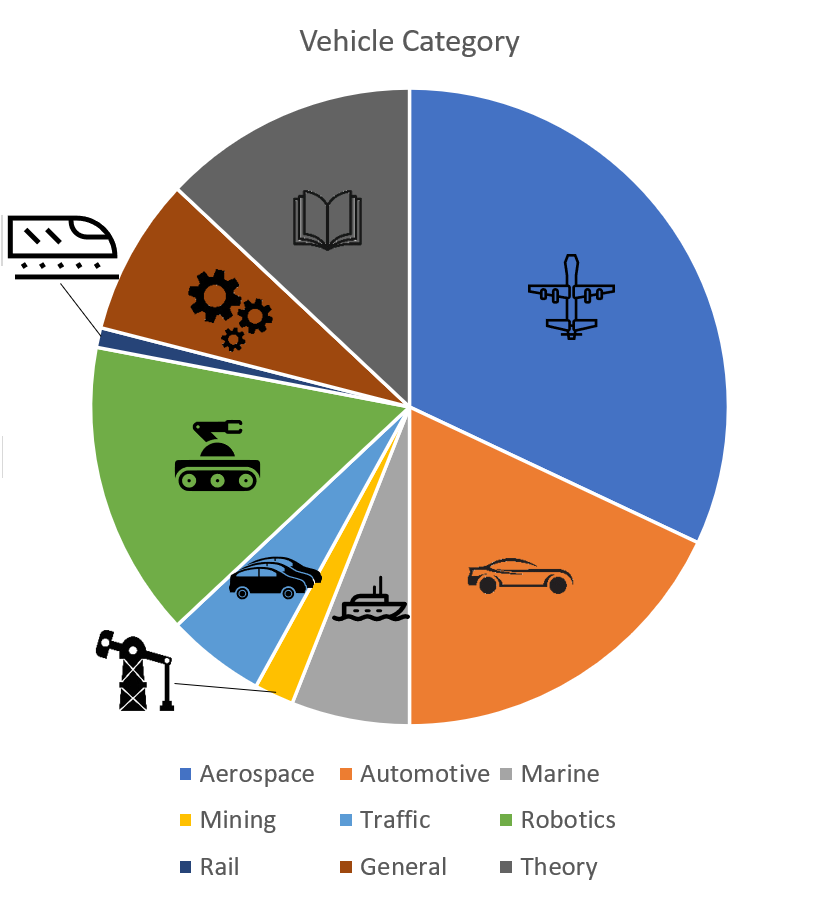}
	\end{minipage}
    \hspace{6ex}
    \begin{minipage}{0.3\textwidth}
    	\vspace{-2ex}
    	\includegraphics[width=0.95\textwidth]{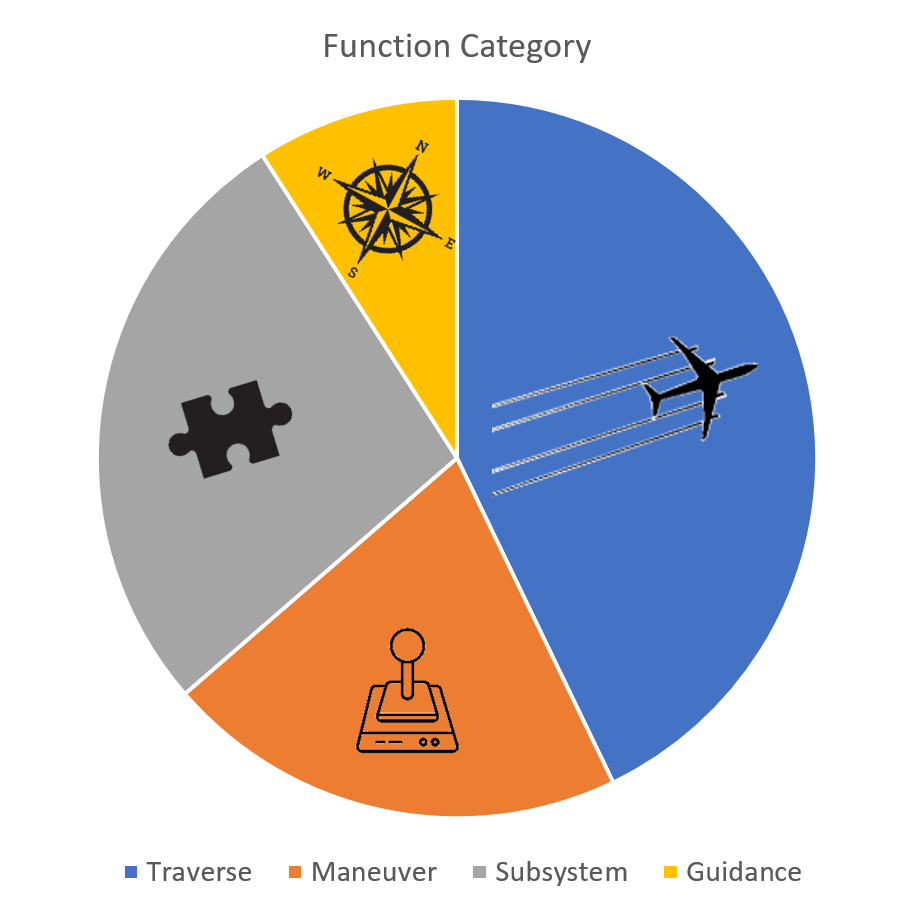}
    \end{minipage}
	\caption{Vehicle categories (left) and function categories (right) of the surveyed smart mobility and 
		vehicular engineering literature. The area of each piece in the pie charts is proportional to the ratio of 
		the number of the conducted studies within each particular area to the number of total studies.}
	\label{fig:VehCat}   
\end{figure*}

Of all the studies found in this survey, Figure \ref{fig:VehCat} (the pie chart on the left) illustrates their proportions in terms of the  type of vehicles represented.  These include aerospace, automotive, marine, mining, traffic, robotics and rail vehicles. Also included are some studies that are theoretical only in terms of presenting a novel Koopman operator-based algorithm or those of general relevance (e.g., pertaining to a generic subsystem or component of multiple possible vehicles) without substantially demonstrating it on any vehicle, whether in simulation, in-vehicle, or on a hardware-in-the-loop test bench.

Figure \ref{fig:VehCat} (the pie chart on the right) further breaks down the proportions of studies that pertain to specific types of functions, rather than vehicle platforms.  These include traverse, maneuver, subsystem and guidance, as defined in what follows. 
 
\begin{itemize}
    \item \textbf{Traverse:} refers to the macro-scale function of a vehicle moving from one point to another (e.g., orbiting). 
    \item \textbf{Maneuver:} refers to a specific mission, operation, reconfiguration or change in situation a vehicle may undertake within its journey (e.g., docking). 
    \item \textbf{Subsystem:} refers to the subject of the study focusing on a component or set of components and their specific operation (e.g., a battery). 
    \item \textbf{Guidance:} refers to the navigational aspect of vehicle path-finding and maintenance, correction or modification of trajectory (e.g., obstacle avoidance). 
\end{itemize}

Finally, ``Traffic Management" is concerned with the coordinated motion of multiple vehicles. Given the uniqueness of certain problems arising in traffic management, we have decided to segregate such studies into their own category.  

\begin{remark}[Machine Learning Community Taxonomy]
With the mainstreaming of Koopman operator-based methods, there also seems to be a linguistic generalization in that the term ``Koopman" or ``Koopman model" is increasingly used to describe any finite state transition matrix approximated for an unmodeled or nonlinear system. This is especially true in the artificial intelligence and machine learning community (see, e.g., \cite{balakrishnan:2021}). Such use may continue to uphold validity since Koopman operator theory remains one of the main formal justifications for utilizing linear state transition matrices for closely capturing the behavior of nonlinear dynamical systems by means of proper linearization. 
\label{rem:MLkoopman}
\end{remark}

\begin{remark}[A Brief Note on CFD studies]
The inclusion of computational fluid dynamics (CFD) studies has generally been avoided in this review. The only exceptions are the studies containing an explicit vehicular application or proposing a new type of Koopman operator-based system identification method/variant. This is because the DMD solution to the Koopman operator was itself first derived in the very context of CFD (see the seminal work by  Schmid~\cite{schmid:2010}) and has since had the most time to mature in the CFD literature. Many such CFD-centric studies involve a generic case study of flow past a cylinder or airfoil, which may have relevance to, e.g., lifting surfaces, screw propellers, or more generically, pumps and turbines.  Thus, the inclusion of literature from the CFD domain poses a vast grey area, often with speculative applicability.  For example, residual DMD (resDMD) was used in a CFD-focused study \cite{colbrook:2022} for supersonic plasma discharge but has significant relevancy to satellite propulsion systems. To narrow the scope of the search, all literature pertaining to the fluid dynamics realm has been excluded, other than those with explicit vehicular applications, or those which identify a novel algorithm (in which case the corresponding study was grouped into the `Theory' category). This takes away a major source of ambiguity, given that much of the pure fluid dynamics literature is generalized (e.g., flow past a cylinder) such that it may or may not be relevant to vehicle motion.  
\label{rem:CFD}
\end{remark}

%

%% file: briefOver.tex
\section{A Brief Overview of Koopman Operator Theory}
\label{sec:briefOver}

Koopman operator theory is founded on the premise that linear transformations of nonlinear dynamical systems exist when represented in Hilbert [function] space \cite{koopman:1931}. This high dimensional space is framed upon a coordinate system consisting of [up to] an infinite number of orthonormal bases (i.e., a linear combination of functions rather than unit vectors), wherein the properties of spatial completeness are preserved. The composition operator (i.e., the ``Koopman" operator) mapping ``observables" between these two spaces could be resolved explicitly as a combination of spectral modes that are related to a dynamical system observed trajectories. Observables can be selected as the system's state and/or some functions thereof. If a set of observables could be found such that the resultant Koopman operator is finite, then those observables form the basis of a ``Koopman-invariant subspace". 

Consider a nonlinear system where the state $\mathbf{x}$ is propagated in time according to 
\begin{equation} 
\mathbf{x}_{k+1}= \mathcal{F} \mathbf{x}_{k},
\end{equation}
\noindent where $\mathbf{x}_k=\mathbf{x}(t_k)$ is the state at time $k$ and $\mathcal{F}$ is a proper dynamical mapping. The premise of the theory is that there exists a Koopman operator, $\mathcal{K}$, which has the property of linearly propagating the observables, $\mathbf{y}\in\mathbb{R}^m$, of any system (including nonlinear and chaotic systems) through Hilbert space \cite{koopman:1931}. In other words, the operator  $\mathcal{K}$ acts according to 
\begin{equation} 
\mathbf{y}_{k+1}=\mathcal{K}\mathbf{y}_{k}, 
\end{equation}
\noindent where $\mathbf{y}_k=\mathbf{y}(t_k)$ is a vector of observables (the state and/or functions thereof) at time $k$, and can be decomposed to a set of observables, $\mathbf{g}$, which may or may not be finite, such that (for brevity of exposition, let us work with the finite presentation)
\begin{equation} \label{eq:observables} 
\mathbf{y}_k=\mathbf{g}(\mathbf{x}_k)=[g_1(\mathbf{x}_k),g_2(\mathbf{x}_k),\dots,g_p(\mathbf{x}_k)]^\top.
\end{equation}
\noindent Additionally, under the property of function composition given by 
\begin{equation} 
\mathcal{K}\left( \mathbf{g}\right)= \mathbf{g}\circ \mathcal{F},
\end{equation}
\noindent ensures that the state transition rule 
\begin{equation} \label{eq:eigenmatrix} 
\mathbf{x}_{k+1}=\mathcal{K}(\mathbf{g}(\mathbf{x}_k))=\mathbf{A}(\mathbf{x}_k),
\end{equation}
%
where $\mathbf{A}:= \mathcal{K}\circ \mathbf{g}$ governs the state propagation through time.

\begin{figure*}[tb]
	\centering
	\includegraphics[width=0.75\textwidth]{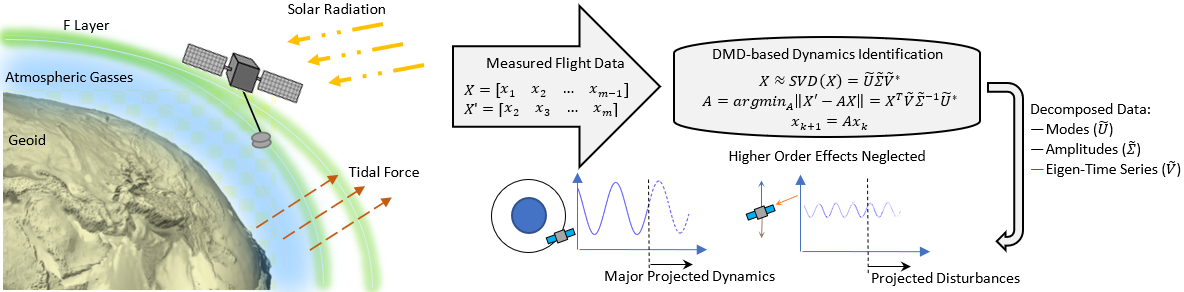}
	\caption{DMD process for linearized, reduced-order model generation of nonlinear Tethered Satellite System subject to unknown disturbances \cite{manzoor:2022b}.}
	\label{fig:DMD_tethered}
\end{figure*}

Historically, determining a Koopman-invariant subspace and computing the matrix  $A$ given by~\eqref{eq:eigenmatrix} was accomplished by trial and error despite being unsuccessful for most dynamical systems \cite{brunton:2016b}. The enabling  engine for  modern-day data-driven applications of the Koopman operator theory are due to the profound insights on geometrical and statistical properties of dynamical systems in the Ph.D. dissertation~\cite{mezic1994geometrical} and the foundational line of work~\cite{mezic2005spectral,mezic2004comparison} on harmonic  analysis of the  Koopman operator.  Mezi\'{c}'s pioneering work~\cite{mezic1994geometrical,mezic2005spectral,mezic2004comparison} along  with   the  computational breakthroughs  based on singular value decomposition (SVD) enabled approximation of the Koopman operator from large amounts of data without relying on the pseudo-inversion of large non-square matrices. It is remarked that \emph{the first modern-day engineering application} of Koopman operator theory has been due to Mezi\'{c} and Banaszuk~\cite{mezic2004comparison} for model parameter identification in combustion rigs.  In what follows, we provide a brief exposition of the main DMD technique that has been the main driving force behind the proliferation of various applications of Koopman operator theory to a plethora of disciplines including geology, epidemiology, finance, and neurology, to name a few (see, e.g., \cite{brunton:2017} and the references therein). Additionally, Figure~\ref{fig:DMD_tethered} provides an intuitive overview of the explained DMD process for the generation of a linearized and reduced-order model of an example nonlinear dynamical system (i.e., a  tethered satellite system subject to unknown disturbances~\cite{manzoor:2022b}). 

From a practical perspective, the matrix $\mathbf{A}$ in~\eqref{eq:eigenmatrix} is the approximation of the Koopman operator acting upon the function space. Since $\mathbf{A}$ is constant in a Koopman-invariant subspace, it may be applied to an entire collection of $m$ measurements, propagating the data matrix $\mathbf{X}$ to the time-shifted data matrix $\mathbf{X}^\prime$, in which the set of observables are arranged column-wise. Specifically, these data matrices are represented as  
\begin{equation} 
\begin{aligned}
&\mathbf{X} = [ \mathbf{x}_0,\cdots,\mathbf{x}_{m-1}], \\
&\mathbf{X}^\prime=\left[\mathbf{x}_1,\cdots,\mathbf{x}_m\right]. 
\end{aligned}
\label{eq:datamat}
\end{equation}

A straightforward and yet computationally  inefficient method for computing an approximation of the Koopman operator can be achieved by multiplying both sides of Equation \ref{eq:eigenmatrix} by the inverse of the data matrix, inv$(\mathbf{X})$. However, this matrix may be too large to invert or non-square. Rather, a more practical solution relies on solving the following optimization problem
\begin{equation} 
\mathbf{A}= \text{argmin}_{\mathbf{A}} \| \mathbf{X}^{\prime} - \mathbf{A}\mathbf{X} \|_F,
\label{eq:optfcn}
\end{equation}
\noindent where $\|\cdot\|_F$ denotes the Frobenius norm.     

To solve the optimization given by~\eqref{eq:optfcn}, regression yields the best-fit fixed Koopman operator, which propagates the selected observables, even if not precisely Koopman-invariant, between any two corresponding columns of the original and time-shifted data matrices. Another way of finding an approximate solution to the minimization problem in Equation~\eqref{eq:optfcn} is to compute proper pseudo-inverses. For instance, SVD-based methods rely on computing the Moore-Penrose left pseudo-inverse. If the data matrix is coincidentally square and invertable, yet the observables are not perfectly Koopman-invariant, then attained solution will not act as a reliable Koopman operator between all sets of corresponding observables. Moreover, a \emph{computational roadblock} exists in that observable data over any practical length of time or collected with a reasonably small sampling time quickly accumulates to a data matrix too large to invert using a desktop computer.   

Methods such as DMD~\cite{schmid:2010} therefore use SVD to obtain a factorization of the transition matrix that is organized by order of modes of decreasing magnitude (see, also, Figure~\ref{fig:DMD_tethered}).  This implies that the major components of the dynamics are captured in a manner that dynamical modes of higher ranks have higher noise-to-signal ratios. Thus, although the dynamics are decomposed into a linear combination of a large set of bases, a reasonable truncation can still be made, which results in a reasonable approximation  for engineering purposes.  One example of this process is illustrated in Figure~\ref{fig:DMD_tethered}, where DMD is used to obtain a linear, reduced-order model of a tethered subsatellite undergoing deployment~\cite{manzoor:2022b} while subjected to multiple environmental disturbances which are too complicated to accurately model, yet whose effects are captured in the observed data. The model truncation capability afforded by the DMD technique and its variants allows for tuning to achieve a tolerable signal-to-noise ratio. 

In the DMD method~\cite{schmid:2010}, the dynamics are decomposed into a linear combination of a large set of bases. Nevertheless, a reasonable truncation, which is suitable for engineering purposes, can be achieved. Essentially, the system dynamics, which are represented by a finite set of nonlinear equations, is approximated with an \emph{up to} infinite set of linear state equations. The order of the obtained linear system can be tuned using a proper reduced-order truncation  method as described later.  Therefore, in the DMD method, we are using the approximated linear system
\begin{equation} \label{eq:statetransition} 
\mathbf{x}_{k+1}=\mathbf{A} \mathbf{x}_k.
\end{equation}

To obtain the operator $\mathbf{A}$ for a general nonlinear system using the SVD-based approach, the snapshots of measurable quantities are obtained from Equation~\eqref{eq:datamat}. For the data matrix $\mathbf{X}$, the following SVD factorization holds
\begin{equation} \label{eq:eigendecompose} 
\mathbf{X}=\mathbf{U}\mathrm{\pmb{\Sigma} \mathbf{V}}.
\end{equation}
\noindent
In the decomposition given by~\eqref{eq:eigendecompose}, $\mathbf{U}$ and $\mathbf{V}$ are unitary matrices and $(\cdot)^\ast$ is the complex conjugate transpose  operator. Moreover, $\pmb{\Sigma}$ is a square matrix of singular values arranged by order of decreasing magnitude, with those in the lower rows corresponding to negligible dynamic modes (i.e., lower signal-to-noise ratio). Thus, the three matrices of the right-hand-side in~\eqref{eq:eigendecompose} can be truncated to rank $r-1$ which maintains the best fit to data. Indeed, $r$ is the optimal hard threshold attained through proper techniques such as the Gavish \& Donoho method (see pp. 31 in \cite{brunton:2019}), to comply with a required truncation size.   

Furthermore, it is possible to obtain the eigendecomposition
\begin{equation} \label{eq:substitution} 
\begin{aligned}
&\mathbf{X} \mathbf{X}^\ast=\mathbf{U}\ \text{diag}(\mathrm{\pmb{\Sigma}}^2,0)\mathbf{U}^\ast \\
&\mathbf{X}^\ast \mathbf{X}=\mathbf{U}\mathrm{\pmb{\Sigma}}^2\mathbf{V}^\ast,
\end{aligned}
\end{equation}
from~\eqref{eq:eigendecompose}. In this eigendecomposition, $\mathbf{U}$ contains the eigenvectors of $\mathbf{X}\mathbf{X}^\ast$ and its columns are ordered according to how much correlation they capture in the columns of $\mathbf{X}$.  A geometric interpretation of the SVD given by~\eqref{eq:substitution} is that it is a product of rotation matrices scaled by the singular values, which is  necessary to project data, $\mathbf{X}$, from the original coordinate system onto a frame wherein the bases of the column-space are defined by $\mathbf{U}$ and the bases of the row-space are defined by $\mathbf{V}$.

Once the data matrix $\mathbf{X}$ from Equation~\eqref{eq:datamat} has been decomposed, the full state transition matrix can be reconstructed according to
\begin{equation} \label{eq:reconstruct} 
\mathbf{A} =\mathbf{X}^{\prime}\widetilde{\mathbf{V}}{\widetilde{\mathrm{\pmb{\Sigma}}}}^{-1}{\widetilde{\mathbf{U}}}^\ast
\end{equation}
\noindent
where $\widetilde{U}$ is also interpreted as the modes of principal orthogonal decomposition and the relationship
\begin{equation}  \label{eq:unitary} 
\widetilde{\mathbf{K}}={\widetilde{\mathbf{U}}}^\ast \mathbf{K}\widetilde{\mathbf{U}}
\end{equation}
holds for the unitary  matrix $\widetilde{\mathbf{U}}$. Finally,  the truncated Koopman matrix can be computed according to
\begin{equation} \label{eq:expansion} 
\widetilde{\mathbf{K}}={\widetilde{\mathbf{U}}}^\ast \mathbf{X}{^\prime}\widetilde{\mathbf{V}}{\widetilde{\mathrm{\pmb{\Sigma}}}}^{-1},   	
\end{equation}
\noindent
where $(\,\,\widetilde\,\,\,)$  denotes the truncated quantities. See the textbook~\cite{brunton:2019} for further details on reverting the  obtained truncated states  back  to the original state space.

Equation \ref{eq:observables} provides  the most general form for choosing the observables. Alternatively, if a catalog of functions (of the states) were included therein, the algorithm would then be referred to as Extended DMD (EDMD). Similarly, in the Sparse Identification of Nonlinear Dynamics (SINDy) algorithm, the time-shifted data matrix (or time derivative of the state, in the continuous time case) is equated to a matrix of possible coefficients projected onto a candidate library of functions to reproduce a structurally linear equivalent system representation  of the nonlinear dynamics. 

The Hankel Alternative View of Koopman (HAVOK) is yet another adaptation of DMD which has a characteristically predictive quality, especially for chaotic systems. This approach relies on Takens embedding theorem, which states that the full dynamics of a chaotic attractor can be reconstructed from the time series of a single measurement diffeomorphic to the original dynamics. We remark that the first instance of utilization of Takens embedding theorem in data-driven Koopman operator theory is due to Mezi\'{c} and Banaszuk~\cite{mezic2004comparison}. This forms a relationship between the Hankel matrix interpretation of all elements propagating through a constant linear transformation of the initial state, and a chaotic system quality of being sensitive to initial conditions. Others have found alternative methods of approximating the Koopman operator (e.g., by use of artificial neural networks), while some have adapted DMD in further creative ways (e.g., Multi-resolution DMD) suited for increased robustness in specific applications. The goal of this paper is to present the application of the Koopman operator (through DMD and its evolved and alternative forms) on applications in the domain of vehicle engineering and smart mobility. The reader is referred to the textbook~\cite{brunton:2019} for more information on Koopman operator theory, SVD, DMD, optimal truncation and other fundamental methods.

%% file: vehicAppli.tex
\section{Literature Review: Vehicular Applications}
\label{sec:vehicApp}

In this section we present our review of  the literature  and categorize the results of over 100 research papers based on both application and algorithm type that are concerned with the applications of Koopman operator theory to the field of smart mobility and vehicular engineering. Table~\ref{tab1} details the specific Koopman operator-based system identification method/algorithm used by the studies referenced hereafter. Given the vast number of algorithms and variants thereof, the reader is encouraged to refer to the respective studies to obtain their technical details. The following presents the surveyed literature organized by vehicle category. 


\subsection{Aerospace}
\subsubsection*{\textbf{Drones/Quadrotors}}   Many of the aerospace applications falling under this review are concerned with \emph{unmanned aerial vehicles} (UAVs), usually of the quadcopter variety.  Specific studies also focus on particular maneuvers, for example, Koopman Eigenfunction Dynamic Mode Decomposition (KEEDMD) has been used for general quadrotor model generation \cite{folkestad:2020b} and, more specifically, to learn the nonlinear ground-effect to improve the speed and quality with which a multi-rotor aircraft may land \cite{folkestad:2020a}.   

Optimization of UAV flight has been explored using Dynamic Mode Decomposition (DMD) \cite{vasisht:2020} and DMD with Control (DMDc) \cite{jin:2020} in optimal control, and by the adjoint Koopman operator \cite{meyers:2019} for expected state propagation, with demonstrated advantages over stochastic control schemes. The adjoint Koopman operator in this literature refers to the left adjoint of the Frobenius-Perron operator.    

Several methods including DMD, Extended DMD (EDMD), bilinear EDMD (biEDMD) and Koopman Canonical Transform have been compared against each other on a planar quadrotor flight testbed, where the superiority of the Koopman Canonical Transform has been demonstrated in handling affine dynamics for nonlinear model predictive control (NMPC).   

Path planning using Robust Koopman Model Predictive Control (RK-MPC) has also been demonstrated in a quadrotor simulation \cite{mamakoukas:2022}. Optimal control for quadcopter stabilization has been demonstrated with models identified using EDMD \cite{berrueta:2020}. Finally, an artificial neural network (ANN)-based approach called Split Koopman Autoencoder has been used in the context of remote state monitoring of UAVs \cite{girgis:2021}, where the communication aspect of flight pertaining to radio frequency signal processing has been addressed.  

\subsubsection*{\textbf{Missiles/Hypersonic Regime}} A few other studies pertaining to the aeronautical domain have also been found to utilize Koopman operator-based methods. This includes the application of \emph{ballistic airdrop}, where the adjoint Koopman operator is used to determine the optimal air release point for ariel delivery to a specified ground target under parametric uncertainty \cite{leonard:2019}. Modeling of \emph{missile dynamics} from noisy data for model predictive control has also been undertaken using Sparse Identification of Nonlinear Dynamics (SINDy) and Stepwise Akaike Information Criteria (SAIC), where it has been shown to be superior  in comparison with state-feedback control \cite{matpan:2021}. 

Remaining the theme of \emph{supersonic flight}, model generation for aerodynamic flutter has been performed using Higher Order DMD (HODMD) to extract frequencies and damping from tests with reduced manual interaction and more robust aeroelastic analysis \cite{le:2019}. Further, in this flow regime, \emph{supersonic combustion ramjet} (ScramJet) engines are susceptible to an ``unstart condition'', which is when the airflow in a duct violently breaks down.  This phenomenon occurs when the pre-combustion shock train (PSCT) location translates upstream beyond the front of the inlet, causing flow separation within the engine and resultant shear layer oscillations. The detection and characterization of this condition relies on accurate modeling of flow characteristics which has been the objective of a study through the use of multi-resolution DMD (mrDMD), demonstrating the inadequacy of the regular DMD approach \cite{sullivan:2021}.  

For \emph{autonomous aircraft} and for \emph{rocket combustion} instability control, principle orthogonal decomposition-based DMD (POD-DMD) was used to simplify equations of motion with a reduced number of variables and selective sensitivity \cite{otto:2022}. POD-DMD is also found to have been used in the computational fluid dynamics (CFD) analysis of flow around an \emph{airfoil} in the sub/transonic regime, with the method increasing computational efficiency by three orders of magnitude while accuracy remained within 5\% as compared to other methods \cite{renganathan:2018}.  

Transitioning between air and space flight, upper and trans-atmospheric dynamics have posed a challenge due to the many environmental factors involved as well as vehicle controllability in what is usually the \emph{hypersonic flight regime}. For this situation, EDMD has been employed to identify a system model for optimal attitude control \cite{mi:2022}.  Similarly, the adjoint Koopman operator has also been used \cite{meyers:2021} to identify equations of motion through a linearly-constrained quadratic program to model atmospheric reentry.  

\subsubsection*{\textbf{Space Systems}} In terms of space system applications, the categories can again be divided into \emph{dynamics-related} (including traversing and maneuvering) and \emph{subsystem-related} (including propulsion). We have found the most application to be the minimum-fuel \emph{orbital rendezvous}.  One approach employed Koopman Map Inversion  to obtain a linearized model for optimal control \cite{servadio:2022a}, while another approach demonstrated Neural Koopman Lyapunov Control for linearizing a generalized affine system \cite{zinage:2022b}. Minimization of the Frobenius norm was performed on a similarly affine \emph{thrust-vectoring} application using the pseudo-inverse to directly solve for the Koopman operator \cite{zinage:2021}.   

In the study \cite{arnas:2022}, a linear model for \emph{zonal harmonics} around the moon was derived using Schur decomposition, rather than a singular value decomposition (SVD) approach to approximate solutions to perturbed ordinary differential equations. A related problem is \emph{lunar station-keeping}, namely for Lyapunov and Halo orbits in the circular-restricted three-body problem (CR3BP).  One study creatively obtained the Koopman operator approximation of the system matrix through direct computation using Legendre polynomials, which are already by their nature a complete and finite set of orthonormal basis in Hilbert space \cite{servadio:2022b}. See \cite{manzoor:2011} for more information on the CR3BP.  

Low thrust trajectory optimization in \emph{underactuated orbital flight} was addressed by a projection method onto vector fields defined by the input matrix \cite{hofmann:2022}. The authors of \cite{manzoor:2022b} demonstrated the extraction of system equations for a tethered subsatellite deployment maneuver subjected to unmodelled dynamics and disturbances using DMD and DMDc. The same objective was achieved by directly using Koopman-invariant observables \cite{chen:2020}, but is not always possible or practical for most problems. Further,  system equations of a \emph{lander} modeled as an inverted pendulum with stabilization thrusters below its center of gravity were derived  using EDMD \cite{bhandari:2021}.  EDMD was also used for a lunar lander in the dynamic allocation of control between a human driver and robot~\cite{broad:2020},~\cite{broad:2018}, referred to as model-based shared control (MbSC).  

On the subsystem side of space applications, DMD was used to find resonant frequencies, damping coefficient and mode shapes in a CFD simulation of a rocket engine's \emph{cryogenic swirl injector} \cite{luong:2021}, the critical flow rate at which vibration occurs, or \emph{``garden hose instability''} (commonly encountered in rocket engines), was investigated using Arnoldi iteration to attain Koopman modes \cite{canham:2017}. 

\begin{remark}[Identified Gap in the Literature]
Although the objective of the aforementioned aerospace-related studies has been limited to system identification, the ultimate goal of almost 37\% of the studies within this vehicle category (which is almost 46\% above all vehicle categories combined together) was to obtain equations of motion in a linear form for the purposes of control using model predictive control (MPC) or other state-space methods. As it can be seen from Fig. \ref{fig:VehCat},  Aerospace has been the largest vehicle class employing emerging Koopman operator-based methods as compared to any other vehicle class.  Despite this, the variety of aerospace vehicle types was found to be quite limited, mostly being small multi-rotor type UAVs.  For this vehicle class, we have not been able to find from the surveyed literature studies pertaining to helicopters and balloons.  
\label{remGap1} 
\end{remark} 


\subsection{Automotive}

\subsubsection*{\textbf{Automobile Engines}} In terms of vehicle subsystems for combustion instability in internal combustion engines (ICE), the Hankel Alternative View of Koopman (HAVOK) method was employed for the prediction of \emph{pre-ignition and super-knock} from real-time peak-pressure data \cite{manzoor:2022a}, while the authors of \cite{ma:2022} employed a portion of the same method (although not by name) to describe the thermoacoustic oscillation characterizing the transition between chaotic states and limit cycles. Also for ICE engines, \emph{turbine dynamics} were investigated (e.g., in superchargers) using EDMD to model and predict turbulent and steady-state behavior \cite{zinage:2022a}.    

\subsubsection*{\textbf{EV Applications}} In terms of \emph{electric vehicles} (EVs), a linear model for motor control was extracted using DMD to actuate a permanent magnet synchronous motor through switching insulated-gate bipolar transistors (IGBTs) \cite{hanke:2018}. IGBTs are a common means to convert direct current (DC) from a battery to the appropriate coils within a motor to control speed in modern EVs. Similarly, an artificial neural network approach has been used to linearize a \emph{DC-DC converter model} for switching control \cite{maksakov:2020}.   Furthermore, linear data-driven predictors afforded by Koopman operator formalism have been utilized to formulate the eco-driving problem for electric vehicles  in a constrained quadratic program setting~\cite{gupta2022koopman},~\cite{shen2022data}. Additionally, data-driven design methods based on Koopman operator theory have been utilized to design X-in-the-loop environments for electrical vehicles~\cite{marco2021data}. Finally, there is a recent body of literature on Koopman operator-based state estimation/prediction and fault diagnosis for batteries that are widely used in electric vehicles~\cite{kanbur2020thermal},~\cite{moreno2023low}.  

\subsubsection*{\textbf{Automotive Model Identification and Control}} Model identification for \emph{nonlinear tire dynamics} using EDMD is investigated in both \cite{cibulka:2019} and \cite{vsvec:2021a}, with the former utilizing a \emph{single-track model} (making it applicable to motorcycles), while the latter further applies MPC control. Similarly, \cite{cibulka:2020} develops an MPC controller with a single-track model obtained by learning Koopman-invariant observables directly from data to recover the vehicle from a nonlinear state (e.g., skidding), when present. MPC is also used to minimize bounce by means of adjusting propulsive force in \cite{buzhardt:2022} while using EDMD for their model generation, thus becoming an alternative method for suspension control. Finally, handling and stability control (with linear time-varying MPC, or LTV-MPC) using \emph{torque vectoring} is explored by \cite{vsvec:2021b} using EDMD for model identification.   

\subsubsection*{\textbf{Autonomous  Vehicle Motion Control/ADAS Systems}} The remaining studies in this vehicle category pertain to motion control in autonomous or \emph{advanced driver-assist systems} (ADAS). In this group, artificial neural network methods were most prevalent.  Control for vehicle motion planning was enabled using Deep Direct Koopman (DDK) or variants in \cite{xiao:2021_i13}, \cite{xiao:2022} and \cite{wang:2021}, with the latter-most specifically applied to a case study dealing with optimal trajectory prediction in racing. Deep learning-based EDMD was employed in \cite{xiao:2022}, also for system identification in path tracking.  For vehicle-to-vehicle related optimized management of traffic comprised of autonomous vehicles, data-driven MPC (DMPC) was employed for coordinated movement \cite{zhan:2022} (e.g., through a controlled intersection) which they term as ``autonomous vehicle platooning''; here their focus is also on the comparison between centralized versus distributed controllers.  The final studies in this category all aim to also obtain a linear model for MPC design and have to do with lane-keeping employing Bilinear Koopman Realization \cite{yu:2022}, Koopman Tracking MPC (KTMPC) \cite{wang:2022}, and WOEDMD \cite{guo:2022} for Operator-AV shared control.   

\begin{remark}[Identified Gap in the Literature]
For this vehicle class, we have not been able to find from the surveyed literature studies pertaining to tracked (including tanks) and screw-propelled vehicles, vehicles otherwise specialized for travel over multiple terrains (e.g., snow, sand, grass, or semi-aquatic environments), as well as tractors, emerging e-mobility devices and other specialized vehicles. Relevant information for applications concerning rovers may be found in the robotics literature, presented in an upcoming section.   
\label{remGap2}  
\end{remark}

\subsection{Marine}

\subsubsection*{\textbf{Autonomous Marine Vehicles}} This vehicle category included some items which could have been categorized instead in the section for robotics, however, where the application dealt specifically with guidance, navigation or propulsion in water, it was considered to be a marine vehicle. This includes a robotic fish, where an EDMD-like algorithm was employed using high-order derivatives of physics-based functions of the state to linearize affine dynamics \cite{sinha:2016}. Similarly, the adjoint Koopman operator was used to improve the efficiency of a \emph{swimming robot} in a flow-like environment \cite{salam:2022}, where it could learn the dynamics of its environment. Finally, a robot was shown to follow a simulated river while avoiding probable locations of unsafe areas (navigation with probabilistic safety constraints) using Naturally Structured DMD (NSDMD)~\cite{pan:2021}, which is actually a modified EDMD algorithm.  

\subsubsection*{\textbf{Oceanic Applications}} The next common theme relates to \emph{oceanic applications}. In the context of an oil spill, oceanic flow was modeled using the adjoint Koopman operator to determine the optimal location for ships to release dispersant to control contaminants in a double-gyre fluid flow field \cite{sinha:2016}. Prediction of wind and oceanic flow patterns was also included in a review that surveyed the use of DMD and its variants \cite{schmid:2022}, including EDMD, Exact DMD, Debiased DMD (also known as forward- and backward-DMD, multiresolution DMD, Hankel DMD (also known as HAVOK), higher-order DMD (which includes derivatives of observable functions) and the adjoint Koopman operator.  However, there are not all applied to vehicular applications, yet is a valuable resource for one who seeks to find an appropriate Koopman operator-based method for a potential vehicular application. Finally, a dissertation by \cite{pan:2021} presents Time-delay DMD (TD-DMD), EDMD, Kernel DMD (KDMD) and Sparsity Promoting DMD (spDMD), and includes the application of model identification for the 3D turbulent air-wake of a ship.  

A unique study has also been found relating to the \emph{measurement of sea ice concentration}, which aims to detect exponentially decaying spatial modes in the Arctic and Antarctic oceans \cite{hogg:2020}. This study is an example of one explicitly relating to satellite data processing, however, there may be many others using Koopman operator-based methods extending into the area of remote sensing and geographic information systems (GIS) which are not within the scope of this survey. 

\begin{remark}[Identified Gap in the Literature]
For this vehicle class, we have not been able to find from the surveyed literature studies pertaining to hovercraft, submarines (including autonomous or remotely piloted underwater vehicles) and offshore platforms. Relevant information for applications concerning hydrofoils may be found in the CFD literature. 
\label{remGap3}  
\end{remark} 

\subsection{Mining}

\subsubsection*{\textbf{Hydraulic Fracturing}} For this category, the most common studies employing Koopman operator-based methods were found for the application of \emph{hydraulic fracturing}, which included \cite{narasingam:2020a}, \cite{klie:2019}, \cite{narasingam:2020b} and \cite{bao:2019}. However, these studies were deemed to fall outside the scope of this review due to their non-vehicular nature.  This is due to the fact that they focused largely on the detection of shale deposits with fixed drilling infrastructure. On the other hand, from a subsystem perspective, it may be somewhat appropriate to include processes enabling natural resource extraction through pipelines.  In that sense, Hankel-based DMD (HDMD) was used to model the \emph{multiphase flow dynamics} of an oil-gas slug and forecast hold-up time profiles \cite{ali:2021}. At the very least, this could have relevance to the operation of inspection/health-monitoring and cleaning vehicles that are typically used in pipelines.  

\subsubsection*{\textbf{Autonomous Excavation}}
Amongst other studies within this category, an \emph{autonomous excavation} application was found where Koopman operator-based system identification was performed using Duel Faceted Linearization for the selection of Koopman invariant observable variables \cite{sotiropoulos:2021}, whereafter an MPC control strategy was applied. Koopman Mode Decomposition (KMD) was used for identifying growing or decaying modes from traffic data and was shown to be superior in performance as compared to artificial intelligence methods \cite{avila:2017,avila2020data}. Finally, the aforementioned study on autonomous vehicle platooning \cite{zhan:2022} can arguably also belong in this category.  

\begin{remark}[Identified Gap in the Literature]
For this vehicle class, we have not been able to find from the surveyed literature studies pertaining to subterranean machines, such as those used for tunnel boring or directional drilling, landships, and elevators. 
\label{remGap4} 
\end{remark} 

\subsection{Traffic}
Traffic management was found to be an area of research where Koopman operator-based system identification techniques are being used. Given its distinctness from physical road vehicles, it has been assigned its own category. The majority of applications in this class of vehicle pertained to \emph{traffic signal phase timing}. In the studies by \cite{ling:2020} and \cite{ling:2018}, DMD was utilized for early identification of unstable queue growth, with the latter further proposing an adaptive traffic control system. The same objective was sought by \cite{lehmberg:2021}, but instead using EDMD to predict pedestrian traffic and an MPC controller for vehicle signaling in response to it.

\subsection{Robotics}

\subsubsection*{\textbf{Robotic Arms}} The operation of \emph{robotic arms} was found to be the most common application in this category and has been treated as a ``vehicle'' for the purposes of this survey as such devices are usually employed to spatially transport a payload from one point to another. For this purpose, EDMD was used (which they refer to as Koopman-MPC) to actuate the arm under voltage disturbance \cite{zhu:2022}. An aforementioned study from the Aerospace category \cite{jin:2020} also demonstrates DMDc and other approaches including ANN and Reinforcement Learning on a robotic arm.  

\subsubsection*{\textbf{Human-Robot Collaboration}} A modified form of EDMD using Autodidact Stiffness Learning was used for detection and adaptivity in applied torque for the human-machine interface of a manipulator (i.e., yoke controller) \cite{goyal:2022}. Similarly, end-effector motion of industrial robotic arms around humans requires environmental state prediction for safe path planning. This was done in one study where the Koopman operator was directly solved for by taking the pseudo-inverse involved in minimization of the Frobenius norm (a computationally expensive operation) \cite{sinha:2022}. The same objective of \emph{safe path planning} was also achieved in \cite{gutow:2020} with the use of the adjoint Koopman operator and in \cite{bujorianu:2021} using the Stochastic Koopman Operator. The latter study also cited an interesting application of their method for automated air traffic management but was not selected for inclusion in the Aerospace category given the lack of demonstration (i.e., simulation, physical experiment or substantive formulation).  

\subsubsection*{\textbf{Soft Robotics}} Other robotics-related applications found in the literature included a \emph{pneumatically actuated soft manipulator} which used EDMD for pick \& place operations for objects of unknown mass \cite{bruder:2021}. Underactuated control of the same type of robot was explored in \cite{haggerty:2020} using Hankel DMD (HDMD). An aforementioned study from the Aerospace category \cite{berrueta:2020} also involved control of a \emph{robotic ball} (called ``Sphero SPRK'') rolling in level sand to follow a predetermined trajectory, EDMD was used here. Similarly, \cite{folkestad:2020b} from the same category also included an example of a wheeled robot using a modified KEEDMD algorithm for mode unknown dynamics and improving computational efficiency.   

\subsubsection*{\textbf{Wheeled/Legged/Swimming Robots}} A unique study involving a wheeled robot utilized EDMD for increased computational efficiency and real-time implementation \cite{shi:2021}. Using \emph{jointed legs} to locomote is another means by which a robot may traverse; the authors of \cite{wu:2021} investigated such crawling and used DMD to expose the method's limitations. Marine robotic applications, as previously mentioned, include the robotic fish \cite{mamakoukas:2021} and obstacle-avoiding river-traversing robot using NSDMD \cite{moyalan:2022}. The former used an MPC controller to make the fish swim in a line or circle, with the linearized model obtained directly from an adaptive error minimization approach using Taylor series error bounds.  

\begin{remark}[Identified Gap in the Literature]
For this vehicle class, we have not been able to find from the surveyed literature studies pertaining to climbing or jumping vehicles, vehicles relying on peristaltic locomotion (see, e.g.,~\cite{Scheraga:2020} for such a robot prototype), attack or surveillance platforms, and robot swarms. Although, a unique biolocomotion study was indeed found to employ DMD to enable mapping between an upper limb and its contra-lateral lower limb while walking forward at constant speed \cite{boudali:2017}. This may arguably qualify as a mode of transportation (i.e., walking), and may very well apply to bipedal robots which are designed to walk like humans. 
\label{remGap5} 
\end{remark} 

\subsection{Rail}
Only one single study was found pertaining to this vehicle category, which was for an MPC application of a \emph{high-speed train} whose linearized model was obtained via EDMD \cite{chen:2021}.  

\begin{remark}[Identified Gap in the Literature]
For this vehicle class, we have not been able to find from the surveyed literature studies pertaining to trams, cable cars and roller coasters. It is important to note that factors surrounding the operation of vehicles or their subsystems were not discounted in the literature search.  For
example, the HAVOK algorithm’s predictive qualities may
 have potential in the areas of environmental forecasting (e.g.,
passenger load, wind and earthquake) and health monitoring
 (e.g., component mean time between failures), such that
vehicles could be operated with appropriate constraints during
 times of expected adverse conditions. 
\label{remGap6} 
\end{remark}

%% file: vehicAppli2.tex
\section{Literature Review: Vehicle-related \& Other Relevant Studies}
\label{sec:vehicApp2}

In this section we provide an overview of other relevant applications, which are not explicitly vehicular in nature, and theoretical/algorithmic variations of the Koopman operator framework that might be beneficial for future applications in the area of smart mobility and vehicular engineering. 

\subsection{General Studies Applicable to Vehicles}
Studies focused on fluid flow are among the most common vehicle-related research topics where Koopman operator theory has played an integral role. Using DMD, the modeling of nonlinear oscillations due to vortex shedding was investigated in \cite{le:2015} and is highly relevant for aeronautical applications.  Also relevant to aeronautical engines that operate in the transonic regime is the use of DMD to model separated and turbulent flow within a convergent-divergent nozzle \cite{leask:2021}, which manifests as the gas path of gas turbine engines. Most internal and external combustion engines also rely on liquid [fuel] injection, for which \cite{leask:2021} is highly relevant as it demonstrates the modeling, prediction and control of nonlinear flow associated with atomization dynamics, enables by SMS and deep convolutional Koopman network (CKN). Similarly, MPC control of nonlinear fluid flow was demonstrated in \cite{bieker:2020} using a deep learning approach. 

The second most common research area found applicable to this category were studies pertaining to motor control, which is especially relevant to UAVs and robots, but potentially also to other types of vehicles when examining them from a subsystem perspective.  This was achieved in \cite{yingzhao:2020} using Gaussian process-based Koopman operator in robust controller design, and in \cite{calderon:2022} using EDMD for current control for the synchronous operation of motors. Finally, a power management study used Stochastic Adversarial Koopman Operator with Auxillary Neural Network for the quick learning of reduced order models that measure the state of charge of Lithium-ion batteries. Potentially applicable to some aircraft and specialized ground vehicles, one study used DMD in the diagnostics of natural gas rotating detonation engines \cite{journell:2020}. 

\subsection{Theoretical Issues with Potential Applications to Smart Mobility and Vehicular Engineering}
\label{subsec:theorIssue}
This section introduces some studies which are theoretically focused on the derivation of unique Koopman operator-based techniques but have not been utilized in the application-focused literature. They are included in this review due to their potential for any future applications the reader may be motivated towards. Firstly, \cite{cohen:2021} uses DMD and rescaled DMD (rDMD) for image processing. Also relating to images, \cite{xiao:2021_i31} used Deterministic and Convolutional Koopman Networks (DCKNet and CKNet, respectively) to predict a suitable trajectory from a provided topography to solve the standard Mountain Car Problem.  This may have relevance to energy-limited adaptive cruise control applications in the automotive category. Linearized, reduced order models in \cite{baddoo:2021} are identified using Physics-informed DMD (piDMD), while \cite{qian:2020} similarly makes a case for physics preservation but utilizes a projection-based model reduction approach. Examples in the former include channel flow and flow past a cylinder, which may be relevant to Marine vehicles, but does not explicitly specify such. The inverted pendulum model is generated in \cite{han:2020} using Deep Neural Network-based Koopman (Koopman DNN), with \cite{bakker:2020} also employing an ANN approach, and \cite{songy:2021} combining ANN with accelerated learning using Deep Koopman Reinforcement Learning (DKRL). Finally, an ANN-based Exact DMD approach is also presented in \cite{li:2021} in comparison with EDMD and LIR-DMD for the development of multi-scale models from coarse data with long-range prediction.  

EDMD and Stochastic Koopman Operator (SKO) are used in \cite{williams:2015}, while \cite{williams:2016} presents a modified EDMD. It has been noted that throughout this survey, EDMD was the most used method in a modified form. Sparse Identification of Nonlinear Dynamics with Control (SINDYc) has been developed in \cite{williams:2015}, and can potentially be applied anywhere DMDc has been used (e.g., in \cite{manzoor:2022b} for tethered subsatellite deployment), although the paper demonstrates its application on a predator-prey model and the Lorenz system. An example in \cite{zhang:2022} applies Robust Tube-based MPC with Koopman (r-KMPC) on the Van der Pol Oscillator, which may have relevance to applications in wireless communication, among other areas. Finally, an interesting application of auto-tuning (i.e., model evaluation) using DMD was presented in \cite{avila:2021} in the context of a zero-sum game.  This may have potential applications in the balancing of parameters and fuzzy criteria in the realm of AV aggressiveness and wargaming.  

%% file: Conclusion.tex

\section{Conclusion}
\label{sec:conclusion}
Since its advent in 1931, Koopman operator theory \cite{koopman:1931} has only recently been actively utilized for solving practical problems, thanks to the introduction of the DMD algorithm in 2008 \cite{schmid:2008}. Since then, a multitude of DMD algorithm variations have risen to prominence and found utility across various fields. A notable feature of our survey paper was reviewing and categorizing the results of over 100 research papers based on both application and algorithm type in smart mobility and vehicle engineering  (see Table~\ref{tab1} and Section~\ref{sec:vehicApp}).  Additionally, this survey paper identified potential research gaps in smart mobility and vehicular engineering applications (Remarks~\ref{remGap1}--\ref{remGap6}). Finally, this review paper discussed theoretical aspects of Koopman operator theory that have been largely neglected by the smart mobility and vehicle engineering community and yet have large potential for contributing to solving open problems in these areas (see Section~\ref{subsec:theorIssue}).

\noindent{\textbf{Future Research Directions.}}	Given the emergence of cyber-threats against connected and autonomous vehicles as well as robotic systems (see, e.g.,~\cite{nekouei2021randomized,mohammadi2022generation}), a future research direction might include utilizing Koopman operator-based algorithms for designing cyber-resilient vehicular and smart mobility applications (see, e.g.,~\cite{taheri2022data} for a related line of research). Another potential research direction is using Koopman operator-based algorithms for predicting the motion of vulnerable road users (VRUs), e.g., pedestrians and cyclists (see, e.g.,~\cite{pool2019context,scholler2020constant}). Finally, rehabilitation robotics and robotic exoskeletons can be the benefactors of the predictive capabilities of Koopman operator-based algorithms for detecting tripping events and/or system  identification in various modes of locomotion (see, e.g.,~\cite{kumar2019extremum,aprigliano2019pre}).

%% file: Koopman_Vehicles_Survey_rev_black.bbl
\begin{thebibliography}{100}
\providecommand{\url}[1]{#1}
\csname url@samestyle\endcsname
\providecommand{\newblock}{\relax}
\providecommand{\bibinfo}[2]{#2}
\providecommand{\BIBentrySTDinterwordspacing}{\spaceskip=0pt\relax}
\providecommand{\BIBentryALTinterwordstretchfactor}{4}
\providecommand{\BIBentryALTinterwordspacing}{\spaceskip=\fontdimen2\font plus
\BIBentryALTinterwordstretchfactor\fontdimen3\font minus
  \fontdimen4\font\relax}
\providecommand{\BIBforeignlanguage}[2]{{%
\expandafter\ifx\csname l@#1\endcsname\relax
\typeout{** WARNING: IEEEtran.bst: No hyphenation pattern has been}%
\typeout{** loaded for the language `#1'. Using the pattern for}%
\typeout{** the default language instead.}%
\else
\language=\csname l@#1\endcsname
\fi
#2}}
\providecommand{\BIBdecl}{\relax}
\BIBdecl

\bibitem{koopman:1931}
B.~O. {K}oopman, ``{H}amiltonian systems and transformation in {H}ilbert
  space,'' \emph{Proceedings of the National Academy of Sciences}, vol.~17,
  no.~5, pp. 315--318, 1931.

\bibitem{mezic1994geometrical}
I.~Mezi{\'c}, \emph{On the geometrical and statistical properties of dynamical
  systems: Theory and applications}.\hskip 1em plus 0.5em minus 0.4em\relax
  California Institute of Technology, 1994.

\bibitem{mezic2005spectral}
------, ``Spectral properties of dynamical systems, model reduction and
  decompositions,'' \emph{Nonlinear Dynamics}, vol.~41, pp. 309--325, 2005.

\bibitem{mezic2004comparison}
I.~Mezi{\'c} and A.~Banaszuk, ``Comparison of systems with complex behavior,''
  \emph{Physica D: Nonlinear Phenomena}, vol. 197, no. 1-2, pp. 101--133, 2004.

\bibitem{schmid:2008}
J.~L. Schmid, Peter J.;~Sesterhenn, ``Dynamic mode decomposition of numerical
  and experimental data,'' in \emph{Bulletin of the American Physical Society},
  ser. 61st APS meeting, vol.~61, San Antonio, 2008, p. 208.

\bibitem{chen:2020}
T.~Chen and J.~Shan, ``{K}oopman-operator-based attitude dynamics and control
  on {SO}(3),'' \emph{Journal of Guidance, Control, and Dynamics}, vol.~43,
  no.~11, pp. 2112--2126, 2020.

\bibitem{cibulka:2020}
V.~Cibulka, T.~Hanis, M.~Korda, and M.~Hromcik, ``Model predictive control of a
  vehicle using {K}oopman operator,'' \emph{IFAC-PapersOnLine}, vol.~53, no.~2,
  pp. 4228--4233, 2020.

\bibitem{hofmann:2022}
C.~Hofmann, S.~Servadio, R.~Linares, F.~Topputo \emph{et~al.}, ``Advances in
  {K}oopman operator theory for optimal control problems in space flight,'' in
  \emph{2022 AAS/AIAA Astrodynamics Specialist Conference}, 2022, pp. 1--13.

\bibitem{qian:2020}
E.~Qian, B.~Kramer, B.~Peherstorfer, and K.~Willcox, ``Lift \& learn:
  Physics-informed machine learning for large-scale nonlinear dynamical
  systems,'' \emph{Physica D: Nonlinear Phenomena}, vol. 406, p. 132401, 2020.

\bibitem{renganathan:2018}
S.~A. Renganathan, Y.~Liu, and D.~N. Mavris, ``{K}oopman-based approach to
  nonintrusive projection-based reduced-order modeling with black-box
  high-fidelity models,'' \emph{AIAA Journal}, vol.~56, no.~10, pp. 4087--4111,
  2018.

\bibitem{otto:2022}
S.~E. Otto, ``Advances in data-driven modeling and sensing for high-dimensional
  nonlinear systems,'' Ph.D. dissertation, Princeton University, 2022.

\bibitem{leonard:2019}
A.~Leonard, J.~Rogers, and A.~Gerlach, ``{K}oopman operator approach to airdrop
  mission planning under uncertainty,'' \emph{Journal of Guidance, Control, and
  Dynamics}, vol.~42, no.~11, pp. 2382--2398, 2019.

\bibitem{schmid:2022}
P.~J. Schmid, ``Dynamic mode decomposition and its variants,'' \emph{Annual
  Review of Fluid Mechanics}, vol.~54, pp. 225--254, 2022.

\bibitem{gutow:2020}
G.~Gutow and J.~D. Rogers, ``{K}oopman operator method for chance-constrained
  motion primitive planning,'' \emph{IEEE Robotics and Automation Letters},
  vol.~5, no.~2, pp. 1572--1578, 2020.

\bibitem{meyers:2019}
J.~J. Meyers, A.~M. Leonard, J.~D. Rogers, and A.~R. Gerlach, ``{K}oopman
  operator approach to optimal control selection under uncertainty,'' in
  \emph{2019 American Control Conference (ACC)}.\hskip 1em plus 0.5em minus
  0.4em\relax IEEE, 2019, pp. 2964--2971.

\bibitem{sinha:2016}
S.~Sinha, U.~Vaidya, and R.~Rajaram, ``Operator theoretic framework for optimal
  placement of sensors and actuators for control of nonequilibrium dynamics,''
  \emph{Journal of Mathematical Analysis and Applications}, vol. 440, no.~2,
  pp. 750--772, 2016.

\bibitem{servadio:2022b}
S.~Servadio, D.~Arnas, and R.~Linares, ``Dynamics near the three-body libration
  points via {K}oopman operator theory,'' \emph{Journal of Guidance, Control,
  and Dynamics}, vol.~45, no.~10, pp. 1800--1814, 2022.

\bibitem{salam:2022}
T.~Salam, V.~Edwards, and M.~A. Hsieh, ``Learning and leveraging features in
  flow-like environments to improve situational awareness,'' \emph{IEEE
  Robotics and Automation Letters}, vol.~7, no.~2, pp. 2071--2078, 2022.

\bibitem{arnas:2022}
D.~Arnas, ``Solving perturbed dynamic systems using {S}chur decomposition,''
  \emph{Journal of Guidance, Control, and Dynamics}, vol.~45, no.~12, pp.
  2211--2228, 2022.

\bibitem{cohen:2021}
I.~Cohen, T.~Berkov, and G.~Gilboa, ``Total-variation mode decomposition,'' in
  \emph{International Conference on Scale Space and Variational Methods in
  Computer Vision}.\hskip 1em plus 0.5em minus 0.4em\relax Springer, 2021, pp.
  52--64.

\bibitem{wu:2021}
Z.~Wu, S.~L. Brunton, and S.~Revzen, ``Challenges in dynamic mode
  decomposition,'' \emph{Journal of the Royal Society Interface}, vol.~18, no.
  185, p. 20210686, 2021.

\bibitem{ling:2020}
E.~Ling, L.~Zheng, L.~J. Ratliff, and S.~Coogan, ``{K}oopman operator
  applications in signalized traffic systems,'' \emph{IEEE Transactions on
  Intelligent Transportation Systems}, vol.~23, no.~4, pp. 3214--3225, 2020.

\bibitem{ling:2018}
E.~Ling, L.~Ratliff, and S.~Coogan, ``{K}oopman operator approach for
  instability detection and mitigation in signalized traffic,'' in \emph{2018
  21st International Conference on Intelligent Transportation Systems
  (ITSC)}.\hskip 1em plus 0.5em minus 0.4em\relax IEEE, 2018, pp. 1297--1302.

\bibitem{boudali:2017}
A.~M. Boudali, P.~J. Sinclair, R.~Smith, and I.~R. Manchester, ``Human
  locomotion analysis: Identifying a dynamic mapping between upper and lower
  limb joints using the {K}oopman operator,'' in \emph{2017 39th Annual
  International Conference of the IEEE Engineering in Medicine and Biology
  Society (EMBC)}.\hskip 1em plus 0.5em minus 0.4em\relax IEEE, 2017, pp.
  1889--1892.

\bibitem{hanke:2018}
S.~Hanke, S.~Peitz, O.~Wallscheid, S.~Klus, J.~B{\"o}cker, and M.~Dellnitz,
  ``{K}oopman operator-based finite-control-set model predictive control for
  electrical drives,'' \emph{arXiv preprint arXiv:1804.00854}, 2018.

\bibitem{le:2015}
S.~Le~Clainche, D.~Rodriguez, V.~Theofilis, and J.~Soria, ``Flow around a
  hemisphere-cylinder at high angle of attack and low reynolds number. {P}art
  {II}: {POD} and {DMD} applied to reduced domains,'' \emph{Aerospace Science
  and Technology}, vol.~44, pp. 88--100, 2015.

\bibitem{vasisht:2020}
S.~Vasisht and M.~Mesbahi, ``Data-guided aerial tracking,'' \emph{Journal of
  Guidance, Control, and Dynamics}, vol.~43, no.~8, pp. 1540--1549, 2020.

\bibitem{journell:2020}
C.~L. Journell, R.~M. Gejji, I.~V. Walters, A.~I. Lemcherfi, C.~D. Slabaugh,
  and J.~B. Stout, ``High-speed diagnostics in a natural gas--air rotating
  detonation engine,'' \emph{Journal of Propulsion and Power}, vol.~36, no.~4,
  pp. 498--507, 2020.

\bibitem{luong:2021}
H.~T. Luong, Y.~Wang, H.-G. Sung, and C.~H. Sohn, ``A comparative study of
  dynamic mode decomposition methods for mode identification in a cryogenic
  swirl injector,'' \emph{Journal of Sound and Vibration}, vol. 503, p. 116108,
  2021.

\bibitem{larusson:2014}
R.~Larusson, N.~Andersson, L.-E. Eriksson, and J.~{\"O}stlund, ``Comparison of
  eigenmode extraction techniques for separated nozzle flows,'' in \emph{50th
  AIAA/ASME/SAE/ASEE Joint Propulsion Conference}, 2014, p. 4003.

\bibitem{avila:2021}
A.~M. Avila, M.~Fonoberova, J.~P. Hespanha, I.~Mezi{\'c}, D.~Clymer,
  J.~Goldstein, M.~A. Pravia, and D.~Javorsek, ``Game balancing using
  {K}oopman-based learning,'' in \emph{2021 American Control Conference
  (ACC)}.\hskip 1em plus 0.5em minus 0.4em\relax IEEE, 2021, pp. 710--717.

\bibitem{manzoor:2022b}
W.~A. Manzoor, S.~Rawashdeh, and A.~Mohammadi, ``{K}oopman operator-based
  data-driven identification of tethered subsatellite deployment dynamics,''
  \emph{ASCE Journal of Aerospace Engineering}, 2023, {A}ccepted, in print.

\bibitem{mamakoukas:2021}
G.~Mamakoukas, M.~L. Castano, X.~Tan, and T.~D. Murphey,
  ``\BIBforeignlanguage{English}{Derivative-based {K}oopman operators for
  real-time control of robotic systems},''
  \emph{\BIBforeignlanguage{English}{IEEE Transactions on Robotics}}, vol.~37,
  no.~6, pp. 2173--2192, 2021.

\bibitem{zinage:2021}
V.~Zinage and E.~Bakolas, ``Far-field minimum-fuel spacecraft rendezvous using
  {K}oopman operator and l2/l1 optimization,'' in \emph{2021 American Control
  Conference (ACC)}.\hskip 1em plus 0.5em minus 0.4em\relax IEEE, 2021, pp.
  2992--2997.

\bibitem{sinha:2022}
A.~Sinha and Y.~Wang, ``{K}oopman operator--based knowledge-guided
  reinforcement learning for safe human--robot interaction,'' \emph{Frontiers
  in Robotics and AI}, vol.~9, 2022, doi:10.3389/frobt.2022.779194.

\bibitem{zhu:2022}
X.~Zhu, C.~Ding, L.~Jia, and Y.~Feng, ``{K}oopman operator based model
  predictive control for trajectory tracking of an omnidirectional mobile
  manipulator,'' \emph{Measurement and Control}, vol.~55, no. 9-10, pp.
  1067--1077, 2022.

\bibitem{berrueta:2020}
T.~A. Berrueta, I.~Abraham, and T.~Murphey,
  \emph{\BIBforeignlanguage{English}{The {K}oopman Operator in Systems and
  Control: Concepts, Methodologies, and Applications}}, A.~Mauroy,
  I.~Mezi{\'c}, and Y.~Susuki, Eds.\hskip 1em plus 0.5em minus 0.4em\relax
  Cham: Springer International Publishing, 2020, vol. 484.

\bibitem{cibulka:2019}
V.~Cibulka, T.~Hanis, and M.~Hromcik, ``Data-driven identification of vehicle
  dynamics using {K}oopman operator,'' \emph{2019 22nd International Conference
  on Process Control (PC19)}, pp. 167--172, 2019.

\bibitem{vsvec:2021a}
M.~{\v{S}}vec, {\v{S}}.~Ile{\v{s}}, and J.~Matu{\v{s}}ko, ``Model predictive
  control of vehicle dynamics based on the {K}oopman operator with extended
  dynamic mode decomposition,'' in \emph{2021 22nd IEEE International
  Conference on Industrial Technology (ICIT)}.\hskip 1em plus 0.5em minus
  0.4em\relax IEEE, 2021, pp. 68--73.

\bibitem{mi:2022}
P.~Mi, Q.~Wu, and Y.~Wang, ``Suboptimal control law for a near-space hypersonic
  vehicle based on {K}oopman operator and algebraic {R}iccati equation,''
  \emph{Proceedings of the Institution of Mechanical Engineers, Part G: Journal
  of Aerospace Engineering}, p. 09544100211045594, 2022.

\bibitem{vsvec:2021b}
M.~{\v{S}}vec, {\v{S}}.~Ile{\v{s}}, and J.~Matu{\v{s}}ko, ``Predictive approach
  to torque vectoring based on the {K}oopman operator,'' in \emph{2021 European
  Control Conference (ECC)}.\hskip 1em plus 0.5em minus 0.4em\relax IEEE, 2021,
  pp. 1341--1346.

\bibitem{zinage:2022a}
S.~Zinage, S.~Jadhav, Y.~Zhou, I.~Bilionis, and P.~Meckl, ``Data driven
  modeling of turbocharger turbine using {K}oopman operator,''
  \emph{IFAC-PapersOnLine}, vol.~55, no.~37, pp. 175--180, 2022.

\bibitem{buzhardt:2022}
J.~Buzhardt and P.~Tallapragada, ``A {K}oopman operator approach for the
  vertical stabilization of an off-road vehicle,'' \emph{IFAC-PapersOnLine},
  vol.~55, no.~37, pp. 675--680, 2022.

\bibitem{lehmberg:2021}
D.~Lehmberg, F.~Dietrich, and G.~K{\"o}ster, ``Modeling {M}elburnians--using
  the {K}oopman operator to gain insight into crowd dynamics,''
  \emph{Transportation Research Part C: Emerging Technologies}, vol. 133, p.
  103437, 2021.

\bibitem{bhandari:2021}
S.~Bhandari, ``Landing a spaceship with {K}oopman operator theory and model
  predictive control,'' Master's thesis, Technical University of Munich, 2021.

\bibitem{li:2021}
M.~Li and L.~Jiang, ``Deep learning nonlinear multiscale dynamic problems using
  {K}oopman operator,'' \emph{Journal of Computational Physics}, vol. 446, p.
  110660, 2021.

\bibitem{williams:2016}
M.~O. Williams, M.~S. Hemati, S.~T. Dawson, I.~G. Kevrekidis, and C.~W. Rowley,
  ``Extending data-driven {K}oopman analysis to actuated systems,''
  \emph{IFAC-PapersOnLine}, vol.~49, no.~18, pp. 704--709, 2016.

\bibitem{calderon:2022}
H.~M. Calder{\'o}n, I.~Hammoud, T.~Oehlschl{\"a}gel, H.~Werner, and R.~Kennel,
  ``Data-driven model predictive current control for synchronous machines: a
  {K}oopman operator approach,'' in \emph{2022 International Symposium on Power
  Electronics, Electrical Drives, Automation and Motion (SPEEDAM)}.\hskip 1em
  plus 0.5em minus 0.4em\relax IEEE, 2022, pp. 942--947.

\bibitem{bruder:2021}
D.~Bruder, X.~Fu, R.~B. Gillespie, C.~D. Remy, and R.~Vasudevan,
  ``{K}oopman-based control of a soft continuum manipulator under variable
  loading conditions,'' \emph{IEEE Robotics and Automation Letters}, vol.~6,
  no.~4, pp. 6852--6859, 2021.

\bibitem{shi:2021}
L.~Shi and K.~Karydis, ``Enhancement for robustness of {K}oopman operator-based
  data-driven mobile robotic systems,'' in \emph{2021 IEEE International
  Conference on Robotics and Automation (ICRA)}.\hskip 1em plus 0.5em minus
  0.4em\relax IEEE, 2021, pp. 2503--2510.

\bibitem{chen:2021}
B.~Chen, Z.~Huang, R.~Zhang, W.~Liu, H.~Li, J.~Wang, Y.~Fan, and J.~Peng,
  ``Data-driven {K}oopman model predictive control for optimal operation of
  high-speed trains,'' \emph{IEEE Access}, vol.~9, pp. 82\,233--82\,248, 2021.

\bibitem{williams:2015}
M.~O. Williams, I.~G. Kevrekidis, and C.~W. Rowley, ``A data--driven
  approximation of the {K}oopman operator: Extending dynamic mode
  decomposition,'' \emph{Journal of Nonlinear Science}, vol.~25, no.~6, pp.
  1307--1346, 2015.

\bibitem{moyalan:2022}
J.~Moyalan, Y.~Chen, and U.~Vaidya, ``Navigation with probabilistic safety
  constraints: A convex formulation,'' in \emph{2022 American Control
  Conference (ACC)}.\hskip 1em plus 0.5em minus 0.4em\relax IEEE, 2022, pp.
  826--830.

\bibitem{broad:2020}
A.~Broad, I.~Abraham, T.~Murphey, and B.~Argall, ``Data-driven {K}oopman
  operators for model-based shared control of human-machine systems,''
  \emph{The International Journal of Robotics Research}, vol.~39, no.~9, pp.
  1178--1195, 2020.

\bibitem{broad:2018}
A.~Broad, T.~Murphey, and B.~Argall, ``Learning models for shared control of
  human-machine systems with unknown dynamics,'' in \emph{Robotics: Science and
  Systems (RSS)}, 2017.

\bibitem{pan:2021}
S.~Pan, ``Robust and interpretable learning for operator-theoretic modeling of
  non-linear dynamics,'' Ph.D. dissertation, University of Michigan, 2021.

\bibitem{sotiropoulos:2021}
F.~E. Sotiropoulos and H.~H. Asada, ``Dynamic modeling of bucket-soil
  interactions using {K}oopman-{DFL} lifting linearization for model predictive
  contouring control of autonomous excavators,'' \emph{IEEE Robotics and
  Automation Letters}, vol.~7, no.~1, pp. 151--158, 2021.

\bibitem{guo:2022}
W.~Guo, H.~Cao, S.~Zhao, M.~Li, B.~Yi, and X.~Song, ``A data-driven model-based
  shared control strategy considering drivers' adaptive behavior in
  driver-automation interaction,'' \emph{Proceedings of the Institution of
  Mechanical Engineers, Part D: Journal of Automobile Engineering}, p.
  09544070221104888, 2022.

\bibitem{yu:2022}
S.~Yu, C.~Shen, and T.~Ersal, ``Autonomous driving using linear model
  predictive control with a {K}oopman operator based bilinear vehicle model,''
  \emph{IFACPapersOnline}, vol.~55, no.~24, pp. 254--259, 2022.

\bibitem{jin:2020}
W.~Jin, Z.~Wang, Z.~Yang, and S.~Mou, ``Pontryagin differentiable programming:
  An end-to-end learning and control framework,'' \emph{Advances in Neural
  Information Processing Systems}, vol.~33, pp. 7979--7992, 2020.

\bibitem{mamakoukas:2022}
G.~Mamakoukas, S.~Di~Cairano, and A.~P. Vinod, ``Robust model predictive
  control with data-driven {K}oopman operators,'' in \emph{2022 American
  Control Conference}.\hskip 1em plus 0.5em minus 0.4em\relax IEEE, 2022, pp.
  3885--3892.

\bibitem{zhang:2022}
X.~Zhang, W.~Pan, R.~Scattolini, S.~Yu, and X.~Xu, ``Robust tube-based model
  predictive control with {K}oopman operators,'' \emph{Automatica}, vol. 137,
  p. 110114, 2022.

\bibitem{folkestad:2020a}
C.~Folkestad, D.~Pastor, and J.~W. Burdick, ``Episodic {K}oopman learning of
  nonlinear robot dynamics with application to fast multirotor landing,'' in
  \emph{2020 IEEE International Conference on Robotics and Automation
  (ICRA)}.\hskip 1em plus 0.5em minus 0.4em\relax IEEE, 2020, pp. 9216--9222.

\bibitem{folkestad:2020b}
C.~Folkestad, Y.~Chen, A.~D. Ames, and J.~W. Burdick, ``Data-driven
  safety-critical control: Synthesizing control barrier functions with
  {K}oopman operators,'' \emph{IEEE Control Systems Letters}, vol.~5, no.~6,
  pp. 2012--2017, 2020.

\bibitem{avila:2017}
A.~M. Avila, ``Applications of {K}oopman operator theory to highway traffic
  dynamics,'' Master's thesis, University of California, Santa Barbara, 2017.

\bibitem{avila2020data}
A.~M. Avila and I.~Mezi{\'c}, ``Data-driven analysis and forecasting of highway
  traffic dynamics,'' \emph{Nature Communications}, vol.~11, no.~1, p. 2090,
  2020.

\bibitem{hogg:2020}
J.~Hogg, M.~Fonoberova, and I.~Mezi{\'c}, ``Exponentially decaying modes and
  long-term prediction of sea ice concentration using {K}oopman mode
  decomposition,'' \emph{Scientific Reports}, vol.~10, no.~1, pp. 1--15, 2020.

\bibitem{canham:2017}
M.~H. Canham, ``On flexible tubes conveying a moving fluid: Variational
  dynamics and spectral analysis,'' Master's thesis, University of Alberta,
  2017.

\bibitem{servadio:2022a}
S.~Servadio, R.~Armellin, and R.~Linares, ``A {K}oopman-operator control
  optimization for relative motion in space,'' in \emph{AIAA SCITECH 2023
  Forum}, 2023, p. 0873.

\bibitem{xiao:2021_i13}
Y.~Xiao, ``{DDK}: A deep {K}oopman approach for dynamics modeling and
  trajectory tracking of autonomous vehicles,'' \emph{arXiv preprint
  arXiv:2110.14700}, 2021.

\bibitem{xiao:2022}
Y.~Xiao, X.~Zhang, X.~Xu, X.~Liu, and J.~Liu, ``Deep neural networks with
  {K}oopman operators for modeling and control of autonomous vehicles,''
  \emph{IEEE Transactions on Intelligent Vehicles}, vol.~8, no.~1, pp.
  135--146, 2023.

\bibitem{wang:2021}
R.~Wang, Y.~Han, and U.~Vaidya, ``Deep {K}oopman data-driven control framework
  for autonomous racing,'' in \emph{Proc. Int. Conf. Robot. Autom.(ICRA)
  Workshop Opportunities Challenges Auton. Racing}, 2021, pp. 1--6.

\bibitem{han:2020}
Y.~Han, W.~Hao, and U.~Vaidya, ``Deep learning of {K}oopman representation for
  control,'' in \emph{2020 59th IEEE Conference on Decision and Control
  (CDC)}.\hskip 1em plus 0.5em minus 0.4em\relax IEEE, 2020, pp. 1890--1895.

\bibitem{wang:2022}
Y.~Wang, Y.~Yang, Y.~Pu, and C.~Manzie, ``Data-driven predictive tracking
  control based on {K}oopman operators,'' \emph{arXiv preprint
  arXiv:2208.12000}, 2022.

\bibitem{bakker:2020}
C.~Bakker, A.~Bhattacharya, S.~Chatterjee, C.~J. Perkins, and M.~R. Oster,
  ``Learning {K}oopman representations for hybrid systems,'' \emph{arXiv
  preprint arXiv:2006.12427}, 2020.

\bibitem{zhan:2022}
J.~Zhan, Z.~Ma, and L.~Zhang, ``Data-driven modeling and distributed predictive
  control of mixed vehicle platoons,'' \emph{IEEE Transactions on Intelligent
  Vehicles}, vol.~8, no.~1, pp. 572--582, 2023.

\bibitem{maksakov:2020}
A.~Maksakov and S.~Palis, ``{K}oopman-based optimal control of boost dc-dc
  converter,'' in \emph{2020 IEEE Problems of Automated Electrodrive. Theory
  and Practice (PAEP)}.\hskip 1em plus 0.5em minus 0.4em\relax IEEE, 2020, pp.
  1--4.

\bibitem{sullivan:2021}
J.~Sullivan, ``Application of data-driven modal decomposition techniques to the
  non-stationary case of scramjet unstart,'' Master's thesis, The Ohio State
  University, 2021.

\bibitem{zinage:2022b}
V.~Zinage and E.~Bakolas, ``Neural {K}oopman {L}yapunov control,''
  \emph{Neurocomputing}, vol. 527, pp. 174--183, 2023.

\bibitem{arbabi2017ergodic}
H.~Arbabi and I.~Mezic, ``Ergodic theory, dynamic mode decomposition, and
  computation of spectral properties of the {K}oopman operator,'' \emph{SIAM
  Journal on Applied Dynamical Systems}, vol.~16, no.~4, pp. 2096--2126, 2017.

\bibitem{ali:2021}
N.~Ali, B.~Viggiano, M.~Tutkun, and R.~B. Cal, ``Forecasting the evolution of
  chaotic dynamics of two-phase slug flow regime,'' \emph{Journal of Petroleum
  Science and Engineering}, vol. 205, p. 108904, 2021.

\bibitem{haggerty:2020}
D.~A. Haggerty, M.~J. Banks, P.~C. Curtis, I.~Mezi{\'c}, and E.~W. Hawkes,
  ``Modeling, reduction, and control of a helically actuated inertial soft
  robotic arm via the {K}oopman operator,'' \emph{arXiv preprint
  arXiv:2011.07939}, 2020.

\bibitem{ma:2022}
Z.~Ma, G.~Wang, T.~Cui, and Y.~Zheng, ``Interpretation of intermittent
  combustion oscillations by a new linearization procedure,'' \emph{Journal of
  Propulsion and Power}, vol.~38, no.~2, pp. 190--199, 2022.

\bibitem{manzoor:2022a}
W.~A. Manzoor, S.~Rawashdeh, and A.~Mohammadi, ``Real-time prediction of
  pre-ignition and super-knock in internal combustion engines,'' \emph{SAE
  International Journal Engines}, vol.~16, no.~3, 2023.

\bibitem{xiao:2021_i31}
Y.~Xiao, X.~Xu, and Q.~Lin, ``Cknet: A convolutional neural network based on
  {K}oopman operator for modeling latent dynamics from pixels,'' \emph{arXiv
  preprint arXiv:2102.10205}, 2021.

\bibitem{leask:2021}
S.~B. Leask, ``Dynamical feature extraction of atomization phenomena using deep
  {K}oopman analysis,'' Ph.D. dissertation, University of California, Irvine,
  2021.

\bibitem{le:2019}
S.~Le~Clainche, R.~Moreno-Ramos, P.~Taylor, and J.~M. Vega, ``New robust method
  to study flight flutter testing,'' \emph{Journal of Aircraft}, vol.~56,
  no.~1, pp. 336--343, 2019.

\bibitem{bieker:2020}
K.~Bieker, S.~Peitz, S.~L. Brunton, J.~N. Kutz, and M.~Dellnitz, ``Deep model
  predictive flow control with limited sensor data and online learning,''
  \emph{Theoretical and Computational Fluid Dynamics}, vol.~34, no.~4, pp.
  577--591, 2020.

\bibitem{castano:2020}
M.~L. Casta{\~n}o, A.~Hess, G.~Mamakoukas, T.~Gao, T.~Murphey, and X.~Tan,
  ``Control-oriented modeling of soft robotic swimmer with {K}oopman
  operators,'' in \emph{2020 IEEE/ASME International Conference on Advanced
  Intelligent Mechatronics (AIM)}.\hskip 1em plus 0.5em minus 0.4em\relax IEEE,
  2020, pp. 1679--1685.

\bibitem{songy:2021}
L.~Songy, J.~Wangy, and J.~Xuz, ``A data-efficient reinforcement learning
  method based on local {K}oopman operators,'' in \emph{20th IEEE International
  Conference on Machine Learning and Applications}.\hskip 1em plus 0.5em minus
  0.4em\relax IEEE, 2021, pp. 515--520.

\bibitem{folkestad:2021}
C.~Folkestad and J.~W. Burdick, ``{K}oopman {NMPC}: {K}oopman-based learning
  and nonlinear model predictive control of control-affine systems,'' in
  \emph{2021 IEEE International Conference on Robotics and Automation}.\hskip
  1em plus 0.5em minus 0.4em\relax IEEE, 2021, pp. 7350--7356.

\bibitem{girgis:2021}
A.~M. Girgis, H.~Seo, J.~Park, M.~Bennis, and J.~Choi, ``Split learning meets
  {K}oopman theory for wireless remote monitoring and prediction,'' in
  \emph{2021 IEEE 32nd Annual International Symposium on Personal, Indoor and
  Mobile Radio Communications (PIMRC)}.\hskip 1em plus 0.5em minus 0.4em\relax
  IEEE, 2021, pp. 1191--1196.

\bibitem{matpan:2021}
H.~Matpan, ``Data driven model discovery and control of longitudinal missile
  dynamics,'' Master's thesis, Middle East Technical University, 2021.

\bibitem{goyal:2022}
T.~Goyal, S.~Hussain, E.~Martinez-Marroquin, N.~A. Brown, and P.~K. Jamwal,
  ``Impedance control of a wrist rehabilitation robot based on autodidact
  stiffness learning,'' \emph{IEEE Transactions on Medical Robotics and
  Bionics}, vol.~4, no.~3, pp. 796--806, 2022.

\bibitem{brunton:2016}
S.~L. Brunton, J.~L. Proctor, and J.~N. Kutz, ``Sparse identification of
  nonlinear dynamics with control ({SINDY}c),'' \emph{IFAC-PapersOnLine},
  vol.~49, no.~18, pp. 710--715, 2016.

\bibitem{balakrishnan:2021}
K.~Balakrishnan and D.~Upadhyay, ``Stochastic adversarial {K}oopman model for
  dynamical systems,'' \emph{arXiv preprint arXiv:2109.05095}, 2021.

\bibitem{yingzhao:2020}
L.~Yingzhao and C.~Jones, ``On {G}aussian process based {K}oopman operators,''
  \emph{IFAC-PapersOnLine}, pp. 52--58, 2020.

\bibitem{bujorianu:2021}
M.~L. Bujorianu, R.~Wisniewski, and E.~Boulougouris, ``Stochastic safety for
  random dynamical systems,'' in \emph{2021 American Control Conference
  (ACC)}.\hskip 1em plus 0.5em minus 0.4em\relax IEEE, 2021, pp. 1340--1345.

\bibitem{chen:2012}
K.~K. Chen, J.~H. Tu, and C.~W. Rowley, ``Variants of dynamic mode
  decomposition: boundary condition, {K}oopman, and {F}ourier analyses,''
  \emph{Journal of Nonlinear Science}, vol.~22, no.~6, pp. 887--915, 2012.

\bibitem{budivsic2012applied}
M.~Budi{\v{s}}i{\'c}, R.~Mohr, and I.~Mezi{\'c}, ``Applied {K}oopmanism,''
  \emph{Chaos: An Interdisciplinary Journal of Nonlinear Science}, vol.~22,
  no.~4, p. 047510, 2012.

\bibitem{mezic2013analysis}
I.~Mezi{\'c}, ``Analysis of fluid flows via spectral properties of the
  {K}oopman operator,'' \emph{Annual Review of Fluid Mechanics}, vol.~45, pp.
  357--378, 2013.

\bibitem{merriamwebster:vehicle}
\BIBentryALTinterwordspacing
Merriam-Webster, ``Vehicle.'' [Online]. Available:
  \url{https://www.merriam-webster.com/dictionary/vehicle}
\BIBentrySTDinterwordspacing

\bibitem{schmid:2010}
P.~J. Schmid, ``Dynamic mode decomposition of numerical and experimental
  data,'' \emph{Journal of Fluid Mechanics}, vol. 656, pp. 5--28, 2010.

\bibitem{colbrook:2022}
M.~J. Colbrook, L.~J. Ayton, and M.~Sz{\H{o}}ke, ``Residual dynamic mode
  decomposition: robust and verified koopmanism,'' \emph{Journal of Fluid
  Mechanics}, vol. 955, p. A21, 2023.

\bibitem{brunton:2016b}
S.~L. Brunton, B.~W. Brunton, J.~L. Proctor, and J.~N. Kutz, ``{K}oopman
  invariant subspaces and finite linear representations of nonlinear dynamical
  systems for control,'' \emph{PLOS One}, vol.~11, no.~2, p. e0150171, 2016.

\bibitem{brunton:2017}
S.~L. Brunton, B.~W. Brunton, J.~L. Proctor, E.~Kaiser, and J.~N. Kutz, ``Chaos
  as an intermittently forced linear system,'' \emph{Nature Communications},
  vol.~8, no.~1, pp. 1--9, 2017.

\bibitem{brunton:2019}
S.~L. Brunton and J.~N. Kutz, \emph{Data-driven science and engineering:
  Machine learning, dynamical systems, and control}.\hskip 1em plus 0.5em minus
  0.4em\relax Cambridge University Press, 2022.

\bibitem{meyers:2021}
J.~Meyers, J.~Rogers, and A.~Gerlach, ``{K}oopman operator method for solution
  of generalized aggregate data inverse problems,'' \emph{Journal of
  Computational Physics}, vol. 428, p. 110082, 2021.

\bibitem{manzoor:2011}
W.~Manzoor, ``A comparative study of formation control at the {E}arth-{M}oon
  {L}2 libration point,'' Master's thesis, Ryerson University, Toronto, ON,
  2011.

\bibitem{gupta2022koopman}
S.~Gupta, D.~Shen, D.~Karbowski, and A.~Rousseau, ``Koopman model predictive
  control for eco-driving of automated vehicles,'' in \emph{2022 American
  Control Conference (ACC)}.\hskip 1em plus 0.5em minus 0.4em\relax IEEE, 2022,
  pp. 2443--2448.

\bibitem{shen2022data}
D.~Shen, J.~Han, D.~Karbowski, and A.~Rousseau, ``Data-driven design of model
  predictive control for powertrain-aware eco-driving considering
  nonlinearities using {K}oopman analysis,'' \emph{IFAC-PapersOnLine}, vol.~55,
  no.~24, pp. 117--122, 2022.

\bibitem{marco2021data}
M.~E. Marco, J.~A. de~la Riva, C.~B. Sopena, and J.~C.~S. Cortes, ``A
  data-driven methodology applied to {X}-in-the-loop environments for electric
  vehicle development,'' in \emph{2021 IEEE Vehicle Power and Propulsion
  Conference (VPPC)}.\hskip 1em plus 0.5em minus 0.4em\relax IEEE, 2021, pp.
  1--5.

\bibitem{kanbur2020thermal}
B.~B. Kanbur, V.~Kumtepeli, and F.~Duan, ``Thermal performance prediction of
  the battery surface via dynamic mode decomposition,'' \emph{Energy}, vol.
  201, p. 117642, 2020.

\bibitem{moreno2023low}
H.~Moreno and A.~Schaum, ``Low-order electrochemical state estimation for
  {L}i-{I}on batteries,'' \emph{Algorithms}, vol.~16, no.~2, p.~73, 2023.

\bibitem{narasingam:2020a}
A.~Narasingam and J.~S.-I. Kwon, ``Application of {K}oopman operator for
  model-based control of fracture propagation and proppant transport in
  hydraulic fracturing operation,'' \emph{Journal of Process Control}, vol.~91,
  pp. 25--36, 2020.

\bibitem{klie:2019}
H.~Klie and H.~Florez, ``Data-driven discovery of unconventional shale
  reservoir dynamics,'' in \emph{SPE Reservoir Simulation Conference}.\hskip
  1em plus 0.5em minus 0.4em\relax OnePetro, 2019, doi:10.2118/193904-MS.

\bibitem{narasingam:2020b}
A.~Narasingam, ``Operator theoretic model predictive control of moving boundary
  dynamical systems: Application to hydraulic fracturing,'' Ph.D. dissertation,
  Texas A\&M University, 2020.

\bibitem{bao:2019}
A.~Bao, E.~Gildin, A.~Narasingam, and J.~S. Kwon, ``Data-driven model reduction
  for coupled flow and geomechanics based on {DMD} methods,'' \emph{Fluids},
  vol.~4, no.~3, p. 138, 2019.

\bibitem{Scheraga:2020}
S.~Scheraga, A.~Mohammadi, T.~Kim, and S.~Baek, ``Design of an underactuated
  peristaltic robot on soft terrain,'' in \emph{2020 IEEE/RSJ International
  Conference on Intelligent Robots and Systems (IROS)}.\hskip 1em plus 0.5em
  minus 0.4em\relax IEEE, 2020, pp. 6419--6426.

\bibitem{baddoo:2021}
P.~Baddoo, B.~Herrmann, B.~McKeon, N.~Kutz, and S.~Brunton, ``Learning and
  enforcing physical structure with physics-informed dynamic mode decomposition
  (pi{DMD}),'' in \emph{APS Division of Fluid Dynamics Meeting Abstracts},
  2021, pp. F10--001.

\bibitem{nekouei2021randomized}
E.~Nekouei, M.~Pirani, H.~Sandberg, and K.~H. Johansson, ``A randomized
  filtering strategy against inference attacks on active steering control
  systems,'' \emph{IEEE Transactions on Information Forensics and Security},
  vol.~17, pp. 16--27, 2021.

\bibitem{mohammadi2022generation}
A.~Mohammadi, H.~Malik, and M.~Abbaszadeh, ``Generation of wheel lockup attacks
  on nonlinear dynamics of vehicle traction,'' in \emph{2022 American Control
  Conference (ACC)}.\hskip 1em plus 0.5em minus 0.4em\relax IEEE, 2022, pp.
  1994--1999.

\bibitem{taheri2022data}
M.~Taheri, K.~Khorasani, N.~Meskin, and I.~Shames, ``Data-driven {K}oopman
  operator based cyber-attacks for nonlinear control affine cyber-physical
  systems,'' in \emph{2022 IEEE 61st Conference on Decision and Control
  (CDC)}.\hskip 1em plus 0.5em minus 0.4em\relax IEEE, 2022, pp. 6769--6775.

\bibitem{pool2019context}
E.~A. Pool, J.~F. Kooij, and D.~M. Gavrila, ``Context-based cyclist path
  prediction using recurrent neural networks,'' in \emph{2019 IEEE Intelligent
  Vehicles Symposium}.\hskip 1em plus 0.5em minus 0.4em\relax IEEE, 2019, pp.
  824--830.

\bibitem{scholler2020constant}
C.~Sch{\"o}ller, V.~Aravantinos, F.~Lay, and A.~Knoll, ``What the constant
  velocity model can teach us about pedestrian motion prediction,'' \emph{IEEE
  Robotics and Automation Letters}, vol.~5, no.~2, pp. 1696--1703, 2020.

\bibitem{kumar2019extremum}
S.~Kumar, A.~Mohammadi, D.~Quintero, S.~Rezazadeh, N.~Gans, and R.~D. Gregg,
  ``Extremum seeking control for model-free auto-tuning of powered prosthetic
  legs,'' \emph{IEEE Transactions on Control Systems Technology}, vol.~28,
  no.~6, pp. 2120--2135, 2020.

\bibitem{aprigliano2019pre}
F.~Aprigliano, S.~Micera, and V.~Monaco, ``Pre-impact detection algorithm to
  identify tripping events using wearable sensors,'' \emph{Sensors}, vol.~19,
  no.~17, p. 3713, 2019.

\end{thebibliography}
